\documentclass[twocolumn]{aastex631}

\newcommand{\paalpha}{Pa$\alpha$} 
\newcommand{\bralpha}{Br$\alpha$} 
\newcommand{\PAHlambda}{3.3 $\mu$m PAH}
\newcommand{\eYSCI}{eYSCI}
\newcommand{\eYSCII}{eYSCII}
\newcommand{\PAHclass}{3.3 $\mu$m PAH peaks}
\def\msun{\hbox{M$_\odot$}}

\received{May  7, 2025}

\begin{document}

\title{FEAST: JWST uncovers the emerging timescales of young star clusters in M83}

\correspondingauthor{Angela Adamo}
\email{angela.adamo@astro.su.se}

\author[0009-0007-2389-0332]{Alice Knutas}
\affiliation{Department of Astronomy, Oskar Klein Centre, Stockholm University, AlbaNova University Center, SE-106 91 Stockholm, Sweden}
\affiliation{Department of Space, Earth and Environment, Chalmers University of Technology, SE-412 93, Göteborg, Sweden.}

\author[0000-0002-8192-8091]{Angela Adamo}
\affiliation{Department of Astronomy, Oskar Klein Centre, Stockholm University, AlbaNova University Center, SE-106 91 Stockholm, Sweden}


\author[0000-0002-8222-8986]{Alex Pedrini}
\affiliation{Department of Astronomy, Oskar Klein Centre, Stockholm University, AlbaNova University Center, SE-106 91 Stockholm, Sweden}

\author[0000-0002-1000-6081]{Sean T. Linden}
\affiliation{Steward Observatory, University of Arizona, 933 N Cherry Avenue, Tucson, AZ 85721, USA}

\author[0009-0008-4009-3391]{Varun Bajaj}
\affiliation{Space Telescope Science Institute, 3700 San Martin Drive Baltimore, MD 21218, USA}

\author[0000-0002-2918-7417]{Jenna E. Ryon}
\affiliation{Space Telescope Science Institute, 3700 San Martin Drive Baltimore, MD 21218, USA}

\author[0000-0003-4910-8939]{Benjamin Gregg}
\affiliation{Department of Astronomy, University of Massachusetts, 710 North Pleasant Street, Amherst, MA 01003, USA}

\author[0000-0001-5189-4022]{Ahmad A. Ali}
\affiliation{I. Physikalisches Institut, Universit\"{a}t zu K\"{o}ln, Z\"{u}lpicher Str. 77, 50937 K\"{o}ln, Germany}

\author[0000-0003-3479-4606]{Eric P. Andersson}
\affiliation{Department of Astrophysics, American Museum of Natural History, 200 Central Park West, New York, NY 10024, USA}

\author[0000-0001-8068-0891]{Arjan Bik}
\affiliation{Department of Astronomy, The Oskar Klein Centre, Stockholm University, AlbaNova University Center, SE-10691 Stockholm, Sweden}

\author[0009-0003-6182-8928]{Giacomo Bortolini}
\affiliation{Department of Astronomy, The Oskar Klein Centre, Stockholm University, AlbaNova University Center, SE-10691 Stockholm, Sweden}

\author[0000-0002-8192-8091]{Anne S.M. Buckner}
\affiliation{Cardiff Hub for Astrophysics Research and Technology (CHART), School of Physics \& Astronomy, Cardiff University, The Parade, CF24 3AA Cardiff, UK}

\author[0000-0002-5189-8004]{Daniela Calzetti}
\affiliation{Department of Astronomy, University of Massachusetts, 710 North Pleasant Street, Amherst, MA 01003, USA}

\author[0000-0002-5259-4774]{Ana Duarte-Cabral}
\affiliation{Cardiff Hub for Astrophysics Research and Technology (CHART), School of Physics \& Astronomy, Cardiff University, The Parade, CF24 3AA Cardiff, UK}

\author[0000-0002-1723-6330]{Bruce~G.~Elmegreen}\affiliation{Katonah, NY 10536, USA}

\author[0000-0002-2199-0977]{Helena Faustino Vieira}
\affiliation{Department of Astronomy, Oskar Klein Center, Stockholm University, AlbaNova University Center, SE-106 91 Stockholm, Sweden}

\author[0000-0001-8608-0408]{John S. Gallagher} 
\affiliation{Dept. Astronomy, U. Wisconsin, 475 North Charter St., Madison, WI 53706 USA}

\author[0000-0002-3247-5321]{Kathryn~Grasha}
\altaffiliation{ARC DECRA Fellow}
\affiliation{Research School of Astronomy and Astrophysics, Australian National University, Canberra, ACT 2611, Australia}   
\affiliation{ARC Centre of Excellence for All Sky Astrophysics in 3 Dimensions (ASTRO 3D), Australia}  

\author[0000-0001-8348-2671]{Kelsey Johnson}\affiliation{Department of Astronomy, University of Virginia, Charlottesville, VA, USA}

\author[0000-0001-8490-6632]{Thomas S.-Y. Lai}
\affil{IPAC, California Institute of Technology, 1200 E. California Blvd., Pasadena, CA 91125}

\author[0009-0009-5509-4706]{Drew Lapeer}
\affiliation{Department of Astronomy, University of Massachusetts, 710 North Pleasant Street, Amherst, MA 01003, USA}

\author[0000-0003-1427-2456]{Matteo Messa}
\affiliation{INAF – OAS, Osservatorio di Astrofisica e Scienza dello Spazio di Bologna, via Gobetti 93/3, I-40129 Bologna, Italy}

\author[0000-0002-3005-1349]{G\"oran \"Ostlin}
\affiliation{Department of Astronomy, Oskar Klein Centre, Stockholm University, AlbaNova University Center, SE-106 91 Stockholm, Sweden}

\author[0000-0003-2954-7643]{Elena Sabbi}
\affiliation{Gemini Observatory/NSFs NOIRLab, 950 N. Cherry Ave., Tucson, AZ 85719, USA} 

\author[0000-0002-0806-168X]{Linda J. Smith}
\affiliation{Space Telescope Science Institute, 3700 San Martin Drive, Baltimore, MD 21218, USA}

\author[0000-0002-0986-4759]{Monica Tosi}
\affiliation{INAF - Osservatorio di Astrofisica e Scienza dello Spazio di Bologna, via Piero Gobetti 93/3, 40129 Bologna, Italy}

\begin{abstract}

We present JWST NIRCam observations of the emerging young star clusters (eYSCs) detected in the nearby spiral galaxy M83. The NIRcam mosaic encompasses the nuclear starburst, the bar, and the inner spiral arms. The eYSCs, detected in Pa$\alpha$ and Br$\alpha$ maps, have been largely missed in previous optical campaigns of young star clusters (YSCs). We distinguish between eYSCI, if they also have compact 3.3~$\mu$m PAH emission associated to them, and eYSCII, if they only appear as compact Pa$\alpha$ emitters. We find that the variations in the 3.3~$\mu$m PAH feature are consistent with an evolutionary sequence where eYSCI evolve into eYSCII and then optical YSCs. This sequence is clear in the F300M-F335M (tracing the excess in the \PAHlambda\ feature) and the F115W-F187N (tracing the excess in Pa$\alpha$) colors which become increasingly bluer as clusters emerge. The central starburst stands out as the region where the most massive eYSCs are currently forming in the galaxy. We estimate that only about 20~\% of the eYSCs will remain detectable as compact YSCs. Combining eYSCs and YSCs ($\leq$10 Myr) we recover an average clearing timescale of 6~Myr in which clusters transition from embedded to fully exposed. We see evidence of shorter emergence timescales ($\sim$5~Myr) for more massive ($>5\times10^3$ \msun) clusters, while star clusters of $\sim 10^3$ \msun\  about 7~Myr. We estimate that eYSCs remain associated to the \PAHlambda\ emission  3--4~Myr. Larger samples of eYSC and YSC populations will provide stronger statistics to further test environmental and cluster mass dependencies on the emergence timescale.

\end{abstract}


\section{Introduction} \label{sec:intro}

Young star clusters (YSCs) in local galaxies are primary tracers carrying information about the early phases and the key mechanisms governing star formation. They are the main birthplaces of massive stars (M~$>$~$8$~M$_\odot$: \citealt{Oey_2004}) and therefore the primary source of stellar feedback: The latter plays a crucial role in regulating the star formation cycle and the evolution of their host galaxy \citep{naabARA&A}.

Most stars form in star clusters, but only some of those clusters are gravitationally bound; the large majority consists of stellar associations that will dissolve within a few crossing times  \citep{portegies2010, krumholz2019}. At early stages, because of the clustered nature of star formation, it remains impossible to distinguish between bound star clusters and associations \citep{adamo2020}, even in the solar neighborhood \citep{bressert2010}. Thus, unless otherwise specified, we will use the term ``star cluster'' for young ($<$~10~Myr) and compact stellar systems of which only a fraction is gravitationally bound and has a higher likelihood to survive in their host galaxies.

Star clusters are formed by the gravitational collapse of a giant molecular cloud (GMC; \citealt{Longmore_2014}). Initially, the collapse produces dense cores evolving into multiple protostars which resemble young stellar objects (YSOs) deeply embedded in their natal cloud \citep{gutermuth2011}. Subsequently, feedback from massive stars ionizes the gas and creates an expanding HII region surrounded by an interface layer, i.e. a photo-dissociation region (PDR) followed by the remaining molecular gas of the GMC \citep{hollenbach_PDR}. During this phase, stars emerge from their natal cloud to become partly embedded and later fully exposed.  The most embedded clusters are typically observed in millimeter, radio and infrared \citep[e.g.,][]{Johnson_2003, Kobulnicky_1999, Reines_2008, Whelan_2011}, while the emerging ones become increasingly visible at shorter wavelengths, for example, near IR (NIR) and optical.  A fully exposed cluster is observed at ultraviolet (UV)-optical wavelengths. In this paper, we will refer to the timescale for a cluster to become fully exposed as the \emph{emerging timescale}. Additionally, we define \emph{emerging young star clusters} (eYSCs) all clusters along this evolutionary phase.

Several studies conducted in Milky Way star-forming regions have revealed that eYSCs are associated with the presence of both H recombination lines, as well as IR emission due to reprocessed far and near UV photons by dust grains and polycyclic aromatic hydrocarbons (PAHs) in PDRs \citep{Churchwell_2009}.  Before the advent of JWST, the study of eYSCs at high resolution in nearby galaxies was mainly performed with H$\alpha$ emission with the Hubble Space Telescope (HST). The H$\alpha$ recombination line represents a tracer of ionized gas powered by OB stars, that dominate the light of eYSCs. Its morphological appearance (from compact to open HII regions) has been used to map the gradual emergence phase of star clusters \citep[e.g.,][]{whitmore2011}. 
Using cluster age-dating obtained from spectral energy distribution (SED) fitting, the emerging timescales appear to last between 4-5~Myr \citep{hollyhead2015studying, Hannon2019}, but could be as fast as 2~Myr \citep{hannon22}, for clusters in the range of M $\sim$ $10^{2.5}-10^{3.5}~$M$_\odot$. 
However, up to 60\% of young star clusters (ages of 1-6~Myr) are missed in optical and UV surveys, and only detected at IR or longer wavelengths \citep{Linden_2023, Messa2021}, making these timescales unreliable. \cite{Messa2021} used the NIR coverage of HST, covering the Pa$\beta$ emission line and found that gas clearing starts after 3~Myr, and is completed by 5~Myr. More recently, \cite{mcquaid2024} and \cite{deshmukh2024} have reported similar results. Overall, these studies are limited to small number statistics (up to several tens of low mass clusters, $\sim1000$~\msun). This raises the question of whether emerging timescales are different for more massive star clusters and/or vary as a function of galactic environment. Since the advent of JWST, we can directly observe embedded star clusters, in particular, the \PAHlambda\ feature has been used to detect embedded massive YSCs \citep[e.g.]{Linden_2023, rodriguez2023, Whitmore_2023}. For example, by cross-matching HST and JWST observations, \citet{Whitmore_2023} reported a completely obscured phase lasting about 1.3~Myr and partially obscured one lasting about 3.7~Myr. In the circumnuclear starburst ring of NGC3351, \cite{Sun_2024} has used JWST, together with observations from Atacama Large Millimeter Array (ALMA) and HST, to detect young massive clusters ($\sim$ $10^5$~M$_\odot$) and divided them into evolutionary stages based on their multi-wavelength properties. By cross-matching multi-wavelength observations, they derived timescales of 2-4~Myr for eYSCs to become visible in NIR, and 4-6~Myr in total to become detectable at optical wavelengths. 

\begin{figure*}
    \centering
    \includegraphics[width=\linewidth]{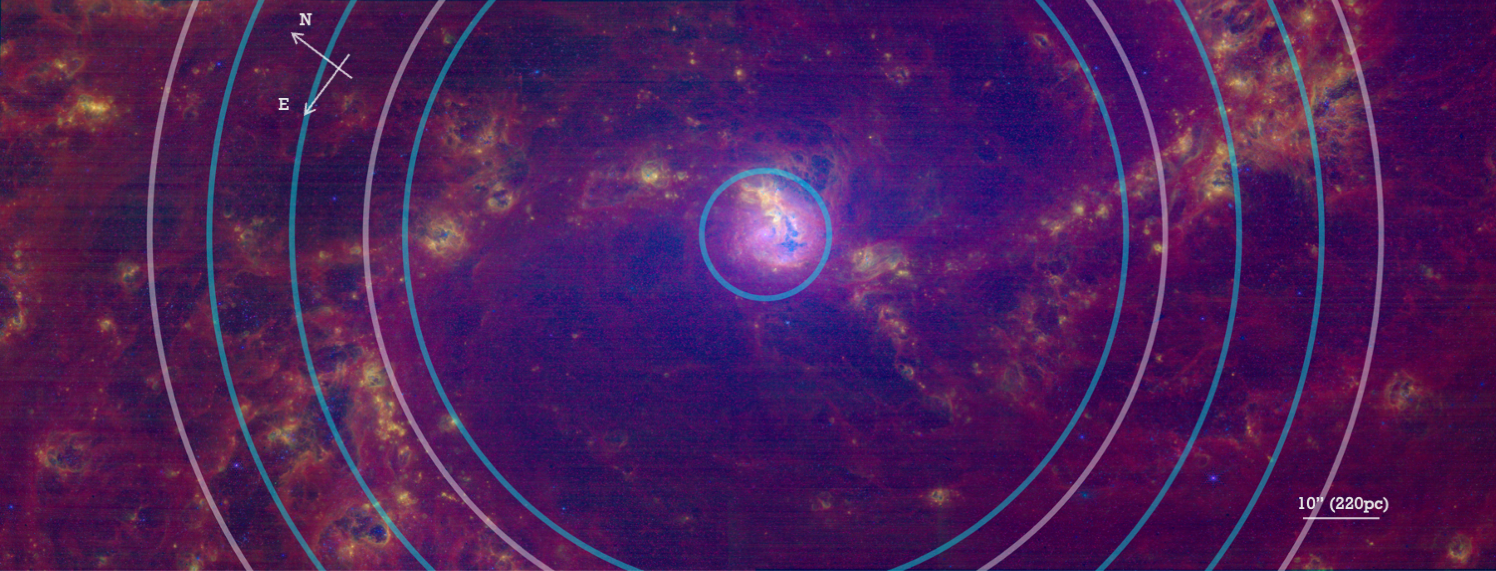}
    \caption{A 3-color NIRCam composite image of the JWST field of view of M83, using the F115W (blue), F187N continum-subtracted (green), and F335M continuum-subtracted (red). The white circles represent informed environmental regions as a function of galactocentric distance: the center, the bar, the end of the bar and outer regions. The light blue circles have been obtained by dividing the galaxy into bins containing same numbers of combined eYSCs and oYSCs. In this second division, we exclude the centre (inner blue circle) and define the bar, the leading inner bar region, the trailing outer bar region and outer region (see Section~\ref{sec: mass_dist_discussion}). }
    \label{fig: enviromental_map}
\end{figure*}

In this work, we will analyse the eYSC population of the nearby star-forming galaxy M83 ($d=4.7$~Mpc; \citealt{Tully+13}) using JWST NIRCam observations obtained under the Feedback in Emerging extragAlactic Star ClusTers  (FEAST, GO 1783, PI Adamo). M83, also known as NGC 5236, is a barred spiral galaxy with $\log_{10}$ M$_*=10.53$~M$_\odot$ \citep{Leroy_2021}. It presents enhanced star formation along the spiral arms and the regions at the end of the bar \citep{adamo2015}. The bar is responsible for feeding the circum-nuclear starburst ring in the center \citep{dellabruna2022galkin} where the most massive star clusters have been detected \citep{harris2001, Wofford_2011}. The FEAST JWST NIRCam mosaic covers roughly 6$\times$2 arcmin, which in physical scale corresponds to about 8.2$\times$2.7~kpc in the North-East direction (see Figure~\ref{fig: enviromental_map}). We sample diverse galactic environments, the nuclear starburst, the bar, end of the bar, spiral arms, including a good portion of the interarm area. In these regions, several works have reported variations in the physical properties of the GMCs and star clusters \citep{freeman2017varying, adamo2015}, as well as in the physical conditions of HII regions \citep{della2022stellar}. In this work we present the observed and physical properties of the complete star cluster population younger than 10 Myr, including eYSCs and estimate the emerging timescale as a function of galactic environment and cluster stellar mass.

The dataset used in this work is presented in Section~\ref{sec: data_and_observations}. In Section~\ref{sec: method} we describe the method used to extract eYSCs. We present the demographics of this newly discovered cluster population and discuss how it relates to the physical properties of optically selected star clusters in Section~\ref{sec: results}. We provide estimates of emerging timescales in different galactic environments and as a function of cluster masses in Section~\ref{sec: discussion}. Our conclusions are presented in Section~\ref{sec: conclusions}.

\section{Observations and data reduction}
\label{sec: data_and_observations}

We use publicly available HST data and newly obtained JWST NIRCam observations of the inner $\sim4$ kpc of the disk of M83. Data reduction of HST and NIRCam observations of the FEAST targets will be presented in detail in Adamo et al.~(in prep.). We provide here a short summary tailored to the observations acquired for M83.

HST observations have been carried out with wide Field Camera 3  instrument (WFC3) under three programs: the WFC3 early release science program GO 11360 (PI Bob O'Connell), GO 12513 (PI Blair) and GO 17225 (PI Calzetti). In this study all data available in the  F275W, F336W, F438W, F547M, F555W, F657N, F689M, and F814W have been downloaded from the MAST archive, reduced and drizzled into mosaics with a common pixel scale of 0.04"/px. The mosaics have been registered to the Gaia reference system.  

The galaxy has been observed in 8  JWST NIRCam bands sampling the stellar continuum and hot dust between 1 and 5 $\mu$m (F115W, F150W, F200W, F300M, F444W), H recombination lines such as Pa$\alpha$ (F187N) and Br$\alpha$ (F405N), and 3.3 $\mu$m PAH feature (F335M). A FULLBOX 4TIGHT primary dither pattern along with a sub-pixel dither pattern (STANDARD with 2 positions) has been used ensuring the coverage of a large FoV$\sim 2.2 \arcmin \times 6\arcmin$ (see Figure~\ref{fig: enviromental_map}) as well as a proper sampling of the NIRCam point spread function (PSF).
The NIRCam data have been reduced using the JWST pipeline and the calibration reference files \texttt{'jwst\_1174.pmap'}. The Gaia registered F814W mosaic has been used as anchor to register the NIRCam F200W mosaic, which in turn has been used as reference to register the remaining NIRCam data. The final NIRCam mosaics have a common pixel scale of 0.04"/pixel and are in units of Jy/pixel.

\subsection{Continuum-subtracted maps} \label{sec: continuum-subtracted_maps}
The JWST F187N, F405N and F335M filters, which are tracers of Pa$\alpha$, Br$\alpha$ and 3.3~$\mu$m PAH emission, respectively, capture a significant fraction of flux coming from stellar continuum and hot dust thermal emission. To account for this fraction, and to obtain final emission maps of these features, we estimated and subtracted the continuum emission from these filters using adjacent broad and medium band filters. We employed an iterative methodology to account for additional emission from these features in the adjacent filters. The reader may find an in-depth description of this process in \cite{gregg24}, as well as a comprehensive discussion on the quality of the resulting maps. In summary, the Pa$\alpha$ emission line map is obtained by subtracting stellar continuum emission using the F150W and the F200W filters. Conversely, we used F300M and F444W to estimate both maps of the 3.3~$\mu$m PAH feature and the Br$\alpha$ emission line.
For each emission map, the detection limit is fixed at 5$\sigma$, where $\sigma$ corresponds to the background root-mean-square value estimated during the point-like sources extraction process (see Section~\ref{sec:optical_catalogue} and Appendix~\ref{app:A}). The 5$\sigma$ limit is equal to $4~\times~10^{-8}$, $2~\times~10^{-8}$ and $1.8~\times~10^{-8}$~Jy/pixel for Pa$\alpha$, Br$\alpha$ and 3.3~$\mu$m~PAH, respectively. For the \paalpha\ and \bralpha\ emission maps the detection limits correspond to an emission rate of hydrogen ionizing photons, a $\log_{10}(Q_0)=47.3$ photons per second. Indicatively, this $Q_0$ value is comparable to the brightness of the emission from an HII region with a circularised radius of 2~pc powered by an O9.5V spectral type star \citep{kennicutt1998star, ISMbook}.

\section{Cluster extraction and photometry} \label{sec: method}
As presented in Adamo et al~(in prep.), we use two different approaches to achieve a complete census of the cluster population from embedded phases to optically bright systems that might survive several hundreds of Myr. This step is necessary as star clusters cover a large color range, making some clusters easier detectable in the NIR (emerging clusters, clusters with red super giants), while others are easier found in optical HST images (e.g. un-obscured clusters).

\subsection{The optically extracted YSC population} \label{sec:optical_catalogue}

We use our newly-developed {\tt FEAST-pipeline} to extract point-like sources in the HST F555W and F547M mosaics as they cover complementary regions of the galaxy (centre and disk). The $V$ band is typically used in HST optical studies of star cluster populations because it is less affected by extinction than UV bands and less prone to contamination by late stages of stellar evolution (e.g, red giants, etc.) affecting the IR bands \citep[e.g.][]{adamo2017}. The pipeline has been described in detail in Adamo et al.~(in prep.), here we describe the main steps taken to produce the photometric catalogs. The extraction step is based on the \texttt{Source extraction and Photometry} (\texttt{SEP}) function \citep{SEP_paper, SEP_paper2}. We use as parameters for the extraction a minimum of 5$\sigma$ detection over 10 contiguous pixels, a background mesh of 30 pixels, a deblending parameter of 32 and contrast 0.0005 (see Table~\ref{tab: source_extr_par}). Since the photometric data are all aligned, we allow an improvement of the centering of each source in the reference filter and use the final position for the next steps. Photometry of all HST and NIRCam bands has been performed using an aperture radius of 5~pixel and a 2 pixel-wide local sky annulus located at a radius of 7~pixel. The concentration index (CI) of each extracted source is measured in the reference band (F555W or F547M) from the magnitude difference at 1 and 3~pixel radius. The measured CI are then used to perform aperture corrections in all the bands.   For the reference filter (F555W or F547M in this case), we extrapolate a relation between the CI of the Moffat model convolved to the filter PSF and the effective radius, R$_{\rm eff}$, of the Moffat model. By construction, the profile of the modeled Moffat has a fixed index of 1.5 and radii changing from 0 to 5~pc. The extrapolated relation between CI vs. R$_{\rm eff}$ is used then to associate to each source a R$_{\rm eff}$ from its CI. This means that for each source, we establish the closest Moffat model that closest describes its light distribution. We then create the shape of the source in all the other bands by convolving this Moffat model to the PSF of each filter. We apply to this model the same criteria used to do aperture photometry and estimate the aperture correction as the difference between the resulting photometry and the total magnitude estimated at 20 pixel (0.8"). 

In total, the initial extraction produces 78427 sources for which photometry and a CI have been estimated. We select among these objects those that are detected with a signal-to-noise higher than 5 ($\sigma_{\sc err}\leq0.2$~mag) in F336W, F438W, F547M or F555W, F814W, and that are clearly distinct from massive single stars: to have an absolute magnitude in V (F547M or F555W) of -6~ABmag (assuming a distance modulus of 28.34 mag), and a CI larger than 1.2 mag (stellar PSF value). This selection results in 13467 sources over the entire FoV. Instead of visually inspect this catalog from scratch to distinguish star cluster candidates from interloopers, we use the latest optical cluster catalog published by \citet{della2022stellar} as a starting point for the visual classification. The latter catalog includes 7459 cluster candidates of which 7280 were published by \citet{adamo2015} and 179 were newly detected in the inner area of radius 0.45~kpc of M83. This catalog contains cluster candidates flagged as ``class 1'' (compact extended sources), which we will refer here as ``class 1$+$2'' following the LEGUS and PHANGS convention \citep{adamo2017, maschmann2024phangs}, and ``class 2'' (multiple peaked, elongated objects) hereafter referred to as ``class 3'' in the convention adopted for the FEAST galaxies). Both the newly extracted FEAST catalog and the \citeauthor{della2022stellar}'s catalog were simultaneously visually inspected.   The majority (7419 out of 7459) of  \citeauthor{della2022stellar}'s cluster candidates were confirmed in this updated inspection, and an additional 455  sources with compact appearance and diffuse light profile in their wings (corresponding to LEGUS\, ``class1$+$2'') were found. In total, this visual inspection has produced 7874 candidates. We repeated the photometry steps for these confirmed candidates allowing improving centering within 1 pixel in x and y direction in the reference band. The final FEAST photometric catalog of the optical clusters contains 7777 star cluster candidates after we apply a 5$\sigma$ selection on the fours HST broadbands (F336W, F438W, F555W or F547M, F814W). This catalog includes also a position flag ``fov'' which is set to 1 if the cluster is within the JWST FoV (see Table~\ref{tab: final_cat_1}).

\subsection{NIR extracted eYSC population} \label{sec: eYSCs_catalogue}

To detect eYSCs we use the emission from the surrounding HII region and PDR as a sign-post for the presence of a star cluster. We start from JWST H-recombination lines: 1.87 $\mu$m \paalpha\ and 4.05 $\mu$m \bralpha\ tracing the HII region and the 3.3 $\mu$m PAH emission map tracing the PDR. The latter emission feature has already been shown to trace embedded star clusters \citep[e.g.][]{rodriguez2023, Linden_2023}.  

The source extraction procedure used for the optical YSCs above is repeated here for the eYSCs, using the {\tt FEAST-pipeline}. We used the extraction parameters listed in Table~\ref{tab: source_extr_par}. In total 7010, 15077 and 11956 sources have been extracted in Pa$\alpha$, Br$\alpha$ and 3.3 $\mu$m PAH maps, respectively. 

To confirm the identified sources, we performed simultaneous visual inspection of the 3 catalogs in the respective emission line maps using SAOImageDs9 \citep[Ds9;][]{ds9_paper} as supporting tool. The goal was to clean each catalogs of residuals, re-center sources, add sources missed in the extraction, and remove sources that did not show a clear peak in emission in their respective emission maps. Before visual inspection, we applied a further selection to the 3.3 $\mu$m PAH extracted sources,  which suffered from a severe contamination of residuals in the continuum--subtracted maps. To simplify the inspection process we applied a magnitude and error cut at 21 mag and 0.1 mag, respectively. This step did not affect the final 3.3 $\mu$m PAH  detections, as we recovered missed compact sources during visual inspection. As a result, the final magnitude distributions in F335M extend to fainter magnitudes than the applied cut.

Next, we used the resulting visual inspected catalogs in \paalpha, Br$\alpha$ and \PAHlambda\ emission to perform photometry using the \texttt{FEAST-pipeline} (described in the previous section). For the eYSC candidates we used the NIRCam F200W as reference filter to estimate the aperture correction. We did not allow further centering to avoid misplacement due to proximity to bright sources. We used an aperture radius  of 5 pixel, and a 2 pixel wide sky annulus of radius 7 pixel. We extract photometry in all HST and JWST bands.  

The sources in the resulting photometric catalogs have been divided into three classes named \eYSCI, \eYSCII\ and \PAHclass, depending on whether their positions in \paalpha\ and 3.3 $\mu$m PAH  overlap. In this step, detection in \bralpha\ was not a necessary condition, because the lower resolution and intrinsically fainter \bralpha\ emission not always allowed us to deblend/detect sources clearly detected in the \paalpha\, map. An overlap is defined when two sources from different catalogs are within a distance of 4 pixels ($\sim$ 3.6 pc). We tested this assumption by allowing instead an overlap of 6 pixels ($\sim$ 5.5 pc), which resulted in only a 2\% increase of \eYSCI\ numbers and 5\% decrease of \eYSCII\ and \PAHclass. We conclude that an adjustment to a larger overlapping distance does not affect the analysis significantly. We use the photometric catalog obtained with \paalpha\ positions to divide \eYSCI\ and \eYSCII\, using the overlap with a \PAHlambda\  peak to identify eYSCI. The 3.3 $\mu$m PAH candidates do not have any clear peaked emission in H recombination lines, and thus we use the photometry obtained with the  3.3 $\mu$m PAH peak positions that do not overlap with any of the H recombination line sources. Within each class, we selected only eYSCs that have detection in NIRCam F187N, F200W, F335M, F405N and F444W, with photometric errors lower than 0.3 mag. The resulting numbers in each class are reported in Table~\ref{tab: final_cat_1}, while coordinates (x, y, right ascension and declination), fluxes (in mJy), magnitudes (AB-system), and concentration indices (CI) are reported in Table \ref{tab: eYSCs_phot_table}.

\begin{deluxetable*}{lcccc} 
\tabletypesize{\scriptsize} 
\tablewidth{0pt} 
\tablecaption{Final number of sources in each eYSC class and optical YSCs after photometric error cuts have been applied. We use this catalog to produce the observe IR colors of the eYSCs in Section~\ref{sec: results}. In the case of the optical YSC catalog, we report between brackets the number of cluster candidates within the NIRCam footprint. In the second row, we list the number of eYSCs and oYSCs after cross-matching their positions, applying a  $\chi_{\text{red}}^2$ $\leq$ 20, and an age cut at 10 Myr in the case of oYSCs. We use these selected clusters in the analysis presented in Section~\ref{sec: SEDfitting_result}. \label{tab: final_cat_1}}
\tablehead{
\colhead{Class} & \colhead{eYSCI }& \colhead{eYSCII} & \colhead{3.3 $\mu$m PAH peaks} & \colhead{YSCs [FOV]}\\
} 
\startdata 
Final photometric catalog & 1126  & 453 & 403 & 7777 [3055] 
      \\ 
Final \# of candidates  & 946  & 361 & 389 & 3195 [829] \\ 
\enddata
 
\end{deluxetable*}

\subsubsection{F200W broadband source extraction} \label{sec: F200W_source_extraction}

\citet{Linden_2023} have proposed an alternative method to select eYSC candidates using only broad and medium band filters in the absence of narrow-band emission line observations. We follow this alternative method here to extract potential eYSCs that are not associated to H recombination line emission. These could potentially be deeply embedded clusters, or low mass systems that have not sampled a massive enough star to power a HII region above our detection limits.
We performed a blind source extraction in the F200W broadband using the \texttt{FEAST-pipeline} with extraction parameters given in Table \ref{tab: source_extr_par} and photometry parameters as those applied to estimate the photometry of the eYSCs presented above. We apply a magnitude cut that selects sources with magnitude error below 0.2 in F335M, F150W, and F200W JWST bands to ensure a high-signal-to-noise. Additionally, to select potentially embedded clusters, we apply a CI cut for the F200W broadband larger than 1.2 mag and the color selection defined as F150W$-$F200W~$>$~0.48 mag  and  F200W$-$F335M~$>$~0 mag \citep{Linden_2023}. A comparison with the already extracted eYSCs shows that 70 eYSC candidates have been potentially missed in our extraction, that is less than 5 \%. We show the recovered IR colors of these sources, referred to as eYSC-BB (eYSCs selected in broadband colors), in Figure~\ref{fig: colorcolorplot_double}. Due to the small numbers and uncertain nature of these sources we have decided to not include them in the SED analysis. 

\subsection{SED-fitting} \label{sec: cigale}

To analyze the SED of our samples of eYSCs and optical YSCs, and to investigate variations in their physical properties, we performed an SED fitting analysis using the Code Investigating GALaxy Emission \citep[CIGALE,][]{cigale_article}. In Linden et al.~(in prep.) and Pedrini et al.~(in prep.), we present in-depth analyses on the CIGALE fitting methodology adopted within the FEAST program, as well as comparisons between different approaches and codes, and the reliability of the recovered physical properties of the star clusters. Here, we present a short summary of the fitting process. 
In a nutshell, CIGALE generates a wide grid of models that characterize the parameter space shaped by stellar, nebular and dust contribution from a source, along with its star formation history. For star clusters, we considered a single burst of star formation with exponential decay factor $\tau_{\rm burst} = 0.001$ Myr. We constrained the lifetime of eYSCs to vary from 1 to 10 Myr, as these objects are selected to be bright in H recombination lines emission. On the other hand, we do not put any age restrictions for the  optical clusters.  The stellar emission grid uses \cite{Bruzual03} single stellar population, with a \cite{Chabrier03} initial mass function (IMF) and solar metallicity, while nebular grids are generated using CLOUDY \citep{Ferland13,cigale_article}. Dust emission models are from \cite{Draine_2014}. We account for dust attenuation using the CIGALE modified starburst model (see \citealt{cigale_article}), which consists of a parametrized version of the starburst attenuation law \citep{Calzetti_2000}. Moreover, the adopted parameter space allows the addition of a reduction factor between the attenuation computed from the emission lines and the stellar continuum attenuation. For the emission lines, we used the \cite{cardelli1989relationship} extinction curve. 
The full grid of models has been fitted for each eYSCs and optical YSCs candidate in our final photometric catalogs (see Table~\ref{tab: final_cat_1}), using HST/WFC3 F225W, F275W, F336W, F438W, F547M, F555W, F657N, F689M, F814W and JWST/NIRCam F115W, F150W, F187N, F200W, F300M, F335M, F405N, F444W. In each filter, a detection lower than 3$\sigma$ has been set as an upper limit. The results of the fitting process include sets of best fitted values for eYSC and optical YSC physical properties (i.e. mass, age and extinction), which are presented in the following section. 

\subsection{Final eYSC and optical YSC catalog in M83} \label{sec: final_SED_catalog}
Because eYSCs and optical YSCs have been extracted independently, we cross-matched the two populations for overlaps using a tolerance of 4 pixels ($\sim$ 3.6 pc) radius. The sources that have entered the catalogs twice have been removed from the eYSCs and tagged as oYSCs. We find that less than 5\% of systems in each eYSC class are in the YSC catalog. We summarize in Table~\ref{tab: final_cat_1} the number of systems in each class. The final photometric catalogs refers to high-confidence level detected  candidates in each class, after magnitude selections have been applied. The final number of candidates in each class refer to objects with reliable SED fit ($\chi_{\text{red}}^2<20$) and uniquely identified in emission line maps or optical broadband colors. For the remaining of this work we will refer to oYSCs optically selected YSCs with ages lower than 10 Myr.

\begin{figure*}
    \centering
    \includegraphics[width=\linewidth]{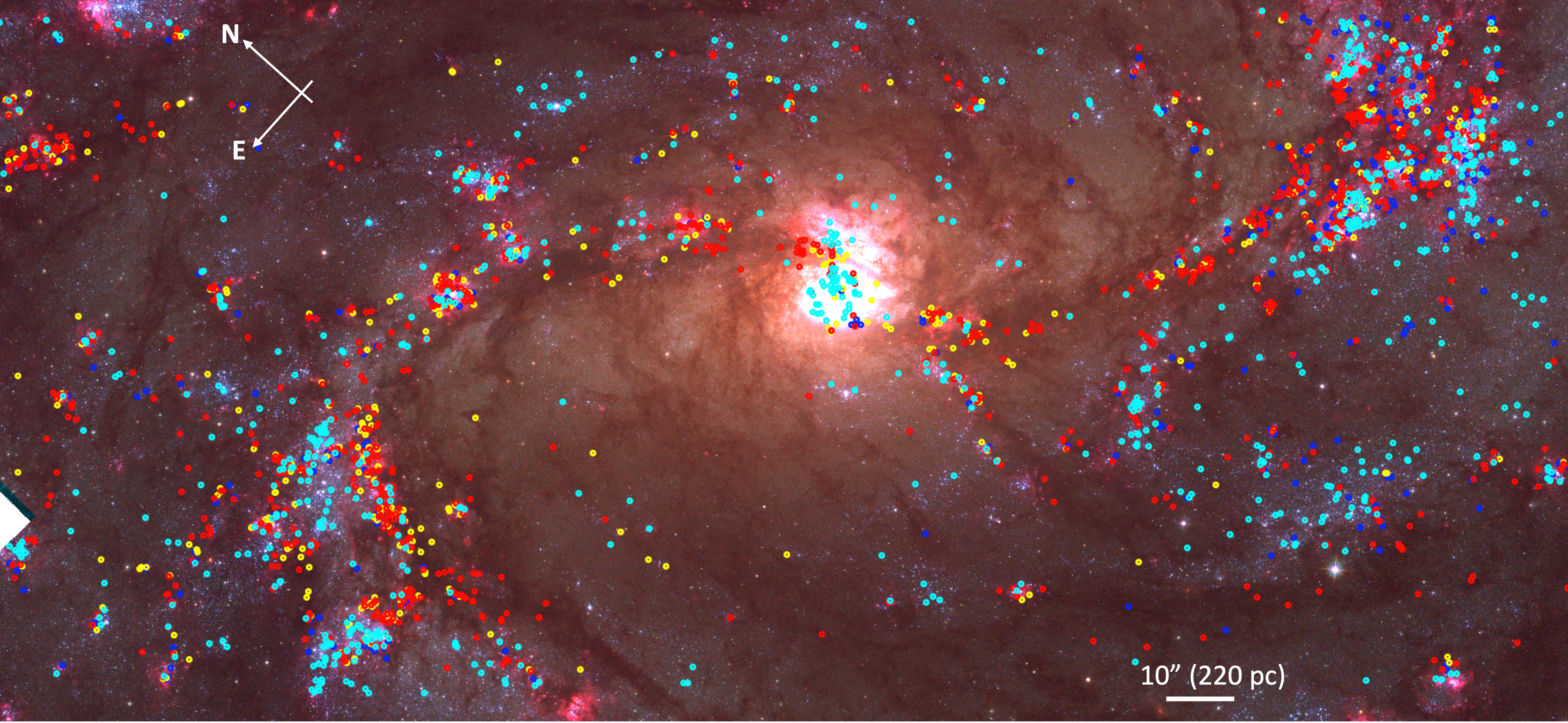}
    \caption{A 3-color HST composite of the FEAST JWST field of view. The mosaic shows the HST/F336W (blue), HST/F438W (green), and HST/F657N (red) bands. Different populations of star clusters are overplotted as circles, with \eYSCI\ in red, \eYSCII\ in yellow, \PAHclass\ in blue and oYSCs (younger than 10 Myr) in cyan. We note that eYSCs are closely located along dusty lanes, while YSC are co-spatial with the regions with the strongest near-UV radiation.}
    \label{fig: M83}
\end{figure*}

We present the identified eYSCs and oYSCs (inside FoV) populations in Figure \ref{fig: M83}. We note that the position of eYSCs coincides with the spiral arms and dust lanes highlighted by the optical colors, while oYSCs are mainly located in UV bright regions, adjacent to spurs and feathers created by the drifting of the spiral arms.

\section{Results} \label{sec: results}

In this section, we explore the observed IR colors of the eYSCs and oYSCs and compare them to the inferred physical properties obtained with CIGALE SED fits.

\subsection{Observed IR colors of emerging YSCs} \label{sec: IRcolors}

\begin{figure*}
    \centering
      \includegraphics[width=0.5\textwidth]{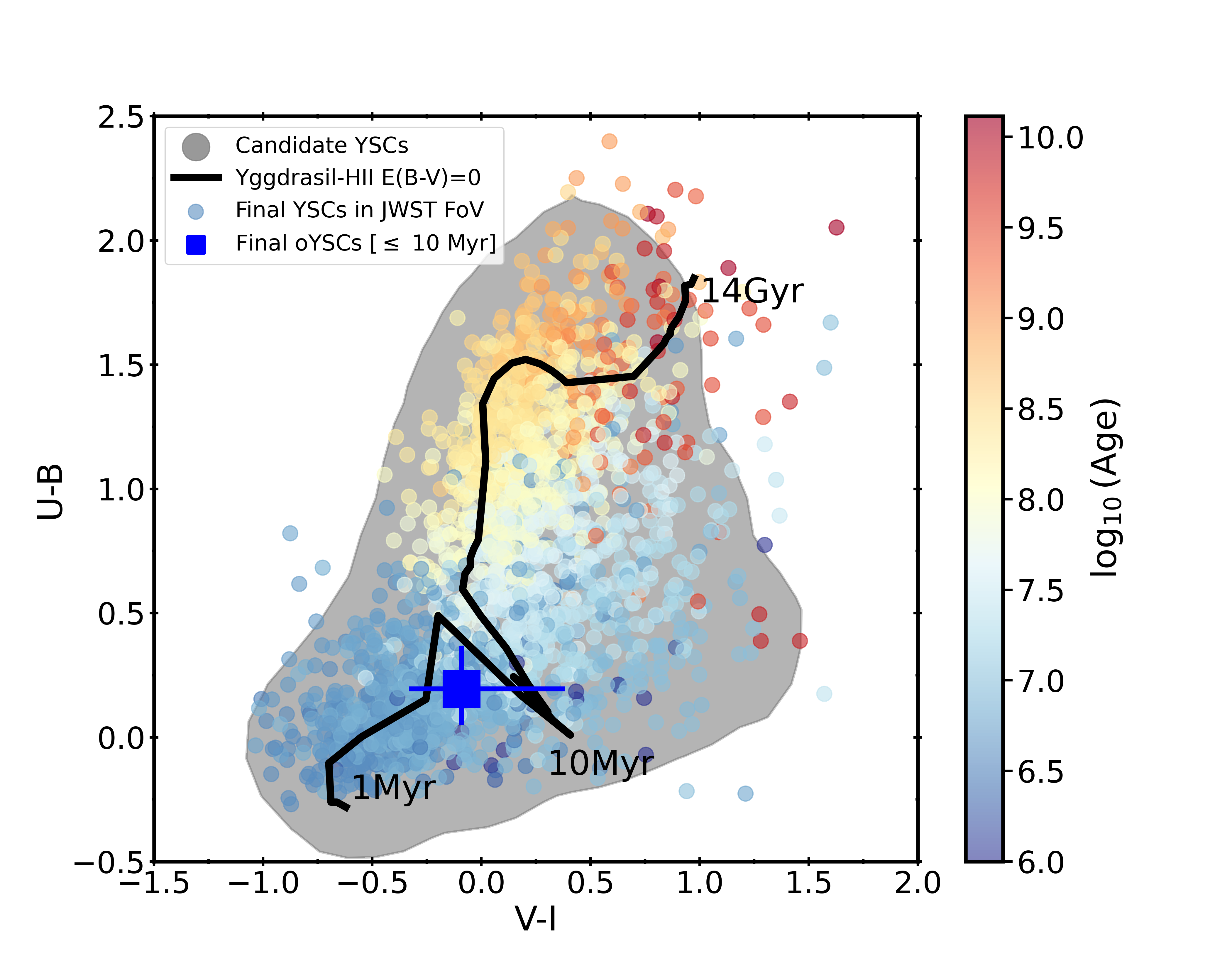}
        \hspace*{-1cm}
            \includegraphics[width=0.5\textwidth]{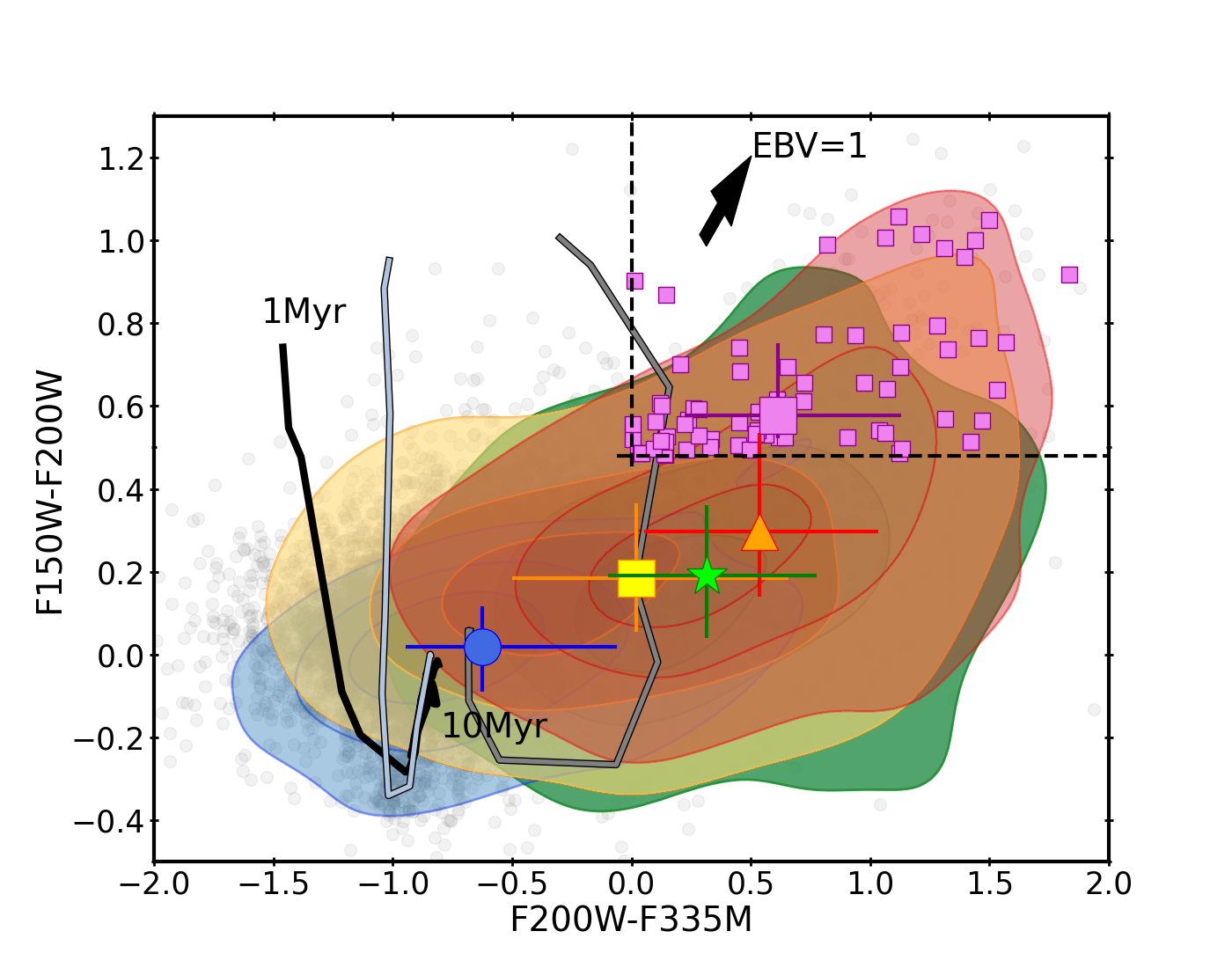}\\
    \includegraphics[width=0.5\textwidth]{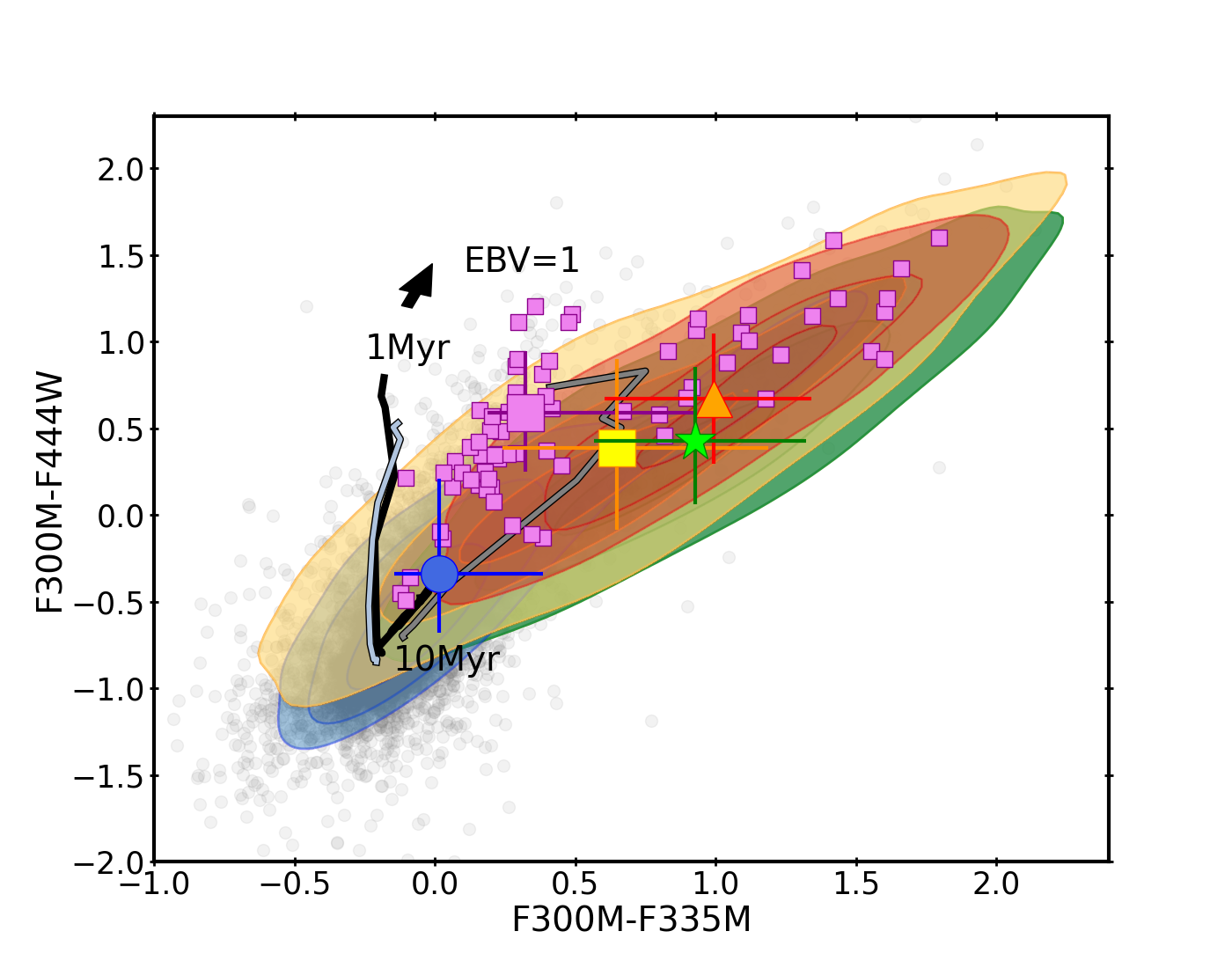}
        \hspace*{-1cm}
     \includegraphics[width=0.5\textwidth]{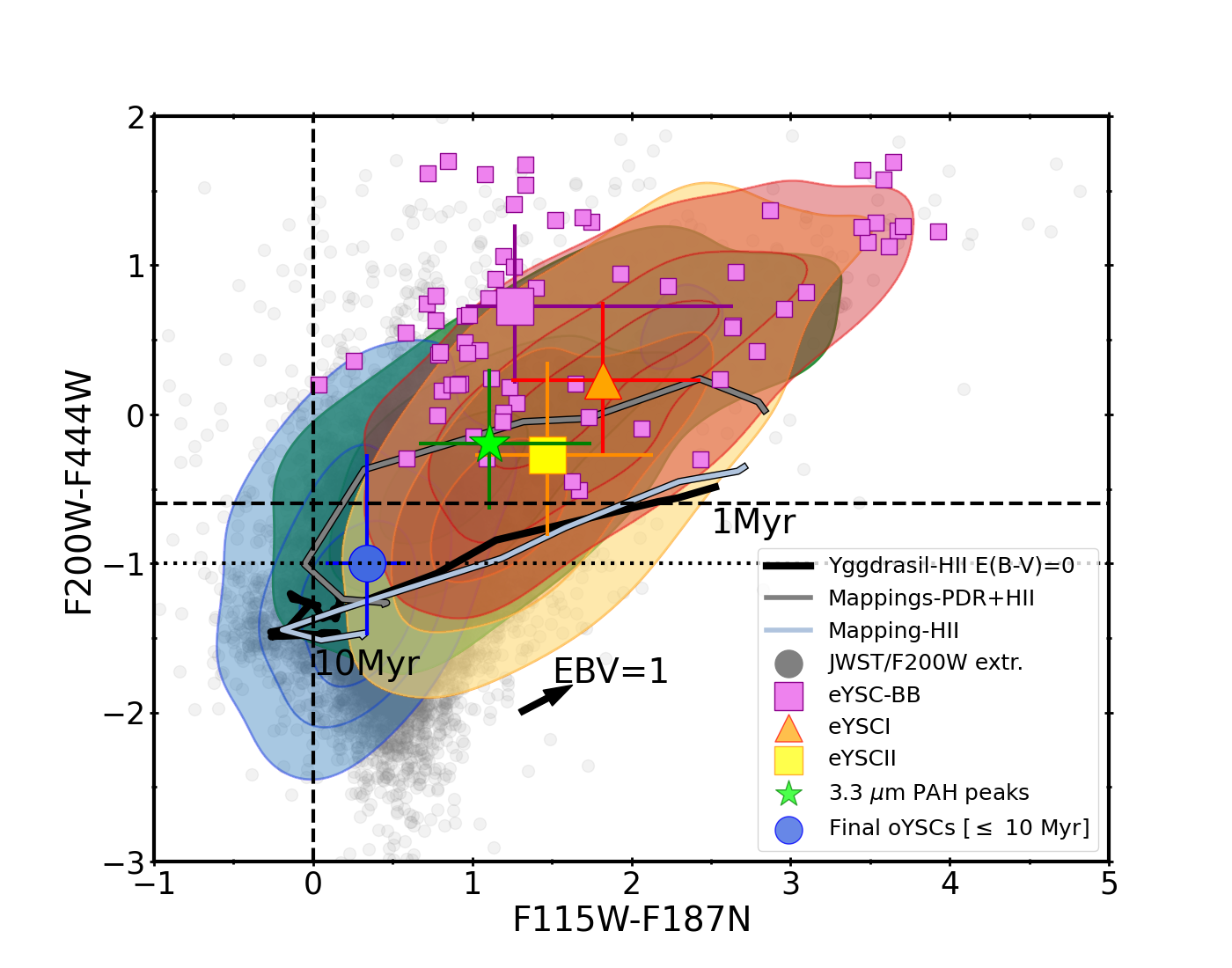} 
    \caption{ Top left: U-B (F336W-F438W) against the V-I (F547M or F555W-F814W) together with a Yggdrasil SPP model (black solid line).  The grey distribution illustrates the entire optical YSC population after visual inspection, while the dots shows the whole YSCs within the field of view (see Table \ref{tab: final_cat_1}) color--coded by the bes--fitted ages. The blue square shows the median colors of  oYSCs ($\leq$ 10 Myr) within the NIRCam FoV and photometric error less than 0.2 mag in F150W, F115W, F187N, F200W, F300M, F335M, and F444W. The NIR color-color diagrams in the upper right, lower left and lower right corner illustrate the  eYSCI (red-orange triangles and red distributions), eYSCII (orange-yellow square and yellow distributions), 3.3 $\mu$m PAH peaks (green star and green distributions), F200W-extracted sources (grey circles), eYSC-BB (purple squares), oYSCs (as defined above, blue circle and blue distributions) together with the Yggdrasil SPP model (black solid lines), Mapping III-HII (light-grey), and Mapping III-PDR+HII models (dark-grey solid lines).  Top right: The F150W-F200W against the F200W-F335M. The box displays the color-color selection for embedded clusters from \cite{Linden_2023}. Lower left: The F300M-F444W against the F300M-F335M color.  Lower Right: The F200W-F444W against the F115W-F187N color together with dashed and dotted lines to illustrate the color space of YSOs.  The distributions are contours with levels at 16\%, 50\% and 84\% percentile. The grey contour in the top left plot shows the 95\% percentile of the total cluster candidates after visual inspection. } 
    \label{fig: colorcolorplot_double} 
\end{figure*} 

We use color-color diagrams as diagnostic diagrams to analyse the cluster population in the galaxy. Color-color plots using UV-optical filters are used to study the properties of the YSCs, while JWST NIR colors are used to describe the observed color of eYSCs that can provide insights into their physical properties. We compare the location of the different cluster populations to two sets of integrated star cluster evolutionary models: Yggdrasil single stellar population (SPP) models \citep{yggdrasilmodel} and Mappings III \citep{mapping}. The Yggdrasil evolutionary track is produced by assuming solar metallicity and sampling the cluster stellar population with a \cite{Kroupa} IMF and  redshifted to $z$ = 0.001711. When relevant, at young ages, the Yggdrasil model includes nebular emission assuming that 50\% of the ionizing photons produced by the SSP ionize the gas in the nebula. On the other hand, the Mappings III model assumes a compactness parameter $C=10^5$, gas pressure $P_0/k=10^5$K/cm$^3$, HI column density $\log$N(HI)$=10^{21.5}$ cm$^{-2}$ and fraction of PDR, where $f_{PDR}=0$ (i.e., no PDR included, labeled MAPPINGS--HII) and $f_{PDR}=1$ (standard PDR model, labeled MAPPINGS-PDR$+$HII). The MAPPINGS model samples ages from 1 to 10 Myr, while YGGDRASIL samples from 1 Myr to 14 Gyr. Model spectra are convolved with the filter responses.

On the top--left side of Figure~\ref{fig: colorcolorplot_double}, we show the U$-$B (F336W$-$F438W) vs. V$-$I (F547M or F555W$-$F814W) colors of the optical YSC population. We show as filled dots the final YSC population (with $\chi_{\text{red}}^2\leq20$ and error lower than 0.2 mag, see Table~\ref{tab: final_cat_1}) within the NIRCam footprint, color-coded accordingly to their best fitted ages from CIGALE. We also highlight the median color of the oYSCs younger than 10 Myr and with a secure detection in the NIRCam filters (photometric error less than 0.2 mag in F150W, F115W, F187N, F200W, F300M, F335M, and F444W). On average, the optical color of the oYSCs is close to the 5-10 Myr parameter space of the Yggdrasil model. We stress that Yggdrasil and CIGALE use the same stellar libraries, thus even if the ages are not derived with the model track used in the plot, their agreement is good (Linden et al.~in prep.). In general, the star cluster population within the FoV is representative of the overall population within the disk, shown by the gray contour. 

In the remaining panels of Figure~\ref{fig: colorcolorplot_double} we focus on the IR colors of the eYSC populations. In the top right plot, we look at the F150W$-$F200W versus the F200W$-$F335M. The y-axis color is sensitive to stellar photosphere, and therefore to stellar evolution. The x-axis color includes the medium band filter F335M which is dominated by the \PAHlambda\, band in the early stages. The MAPPING models that include treatment of this feature show indeed a large variation in the latter color during the early stages of cluster evolution. We also include a color excess arrow illustrating the direction of increased extinction, estimated with the Extinction Python package \citep{extinctionlaw} assuming a \cite{cardelli1989relationship} extinction law and the gradient of the extinction curve at visible wavelength, $R_V=3.1$ \citep{ISMbook}. \citet{Linden_2023} used this color space to isolate the most embedded star clusters in massive starburst galaxies. Their proposed color selection, i.e., the top right area enclosed by the dashed lines, is based on the presence of extinction and excess emission in the \PAHlambda\, band, and overlaps with the color distribution of the reddest eYSCs. By focusing on the median colors, we see a clear evolution sequence from eYSCI (red-orange triangles), to eYSCII (orange-yellow squares), and oYSCs (blue circle), suggesting the emergence of the star clusters and the rapid disappearance of the PDR. After applying the color selection \citep{Linden_2023} to our catalog of eYSCs, we find  319 (28.0\%)  eYSCI,  64 (14.0\%)  eYSCII,  and 61 (15.0\%) 3.3 $\mu$m PAH peaks occupy the upper right corner area. These are the fractions of eYSCs that have comparable colors to the embedded star clusters as defined and selected by Linden and co-authors.

In the lower left plot, we show another color combination. On the y-axis we plot the F300M-F444W with the F300M still dominated by stellar emission. The wide-band  F444W filter is dominated by several molecular and H-line emission lines as well as a hot dust component, especially strong in star-forming regions. The F300M-F335M is another indicator of the strength of the PAH band.  On average eYSCs spread again in an  evolutionary sequence with the oYSCs sitting closest to the star clusters models when the F444W in included.

In the lower right plot, we compare the F200W-F444W versus the F115W-F187N. The color used in the y-axis has been used (at similar wavelengths) in studies of YSOs in the Magellanic Clouds with JWST \citep{habel2024} and in the massive star-forming region Cygnus-X with Spitzer \citep{pokhrel2020} and is sensitive to the hot dust. The x-axis color is sensitive to the \paalpha\ excess. On average, we see that eYSCs occupy the same color space of YSOs in galaxies in the Local Volume, shown by the dashed \citep[F115W$-$F187N~$>$~0 and  F200W$-$F444W~$>$~-0.6,][]{habel2024} and and dotted lines  \citep[F200W$-$F444W~$>$~-0.6][]{pokhrel2020}. Their average \paalpha\ emission is significantly stronger than in the oYSCs, even though the latter have still quite young ages. Overall the behavior of the oYSCs is expected if their average ages are older than the eYSCs.  

The picture conveyed by the NIR color analysis is that with our emission-based extraction we are able to detect recently formed star clusters in a broad range of emergence states. On average, the eYSCs are characterized by red 4-$\mu$m colors, and strong emission in hydrogen and \PAHlambda\ emission. Their  4-$\mu$m colors fully overlap with YSOs observed within the Local Volume, suggesting that a significant fraction of their stars might have not yet reached the main sequence and are in the accretion phase. On the other hand, the oYSCs show a spread, and  we clearly see that a fraction of the optically detected clusters with ages less  than 10 Myr overlap with the space occupied by eYSCs, but their NIR properties are less extreme, as already reported by \citet{rodriguez2024}.

\subsection{Clusters physical properties from the SED fitting analysis} \label{sec: SEDfitting_result}

Next, we present the distributions of the physical properties derived from the SED-fitting analysis described in Section~\ref{sec: cigale}. We focus on eYSCs and oYSCs candidates that have solid detections and good $\chi_{\text{red}}^2$ (see Table \ref{tab: final_cat_1}). For the oYSCs, we select only those within the NIRCam FoV. We will first describe the recovered trends in cluster ages, masses, and extinctions, and then discuss their reliability.  

\begin{figure*}

    \includegraphics[width=\linewidth]{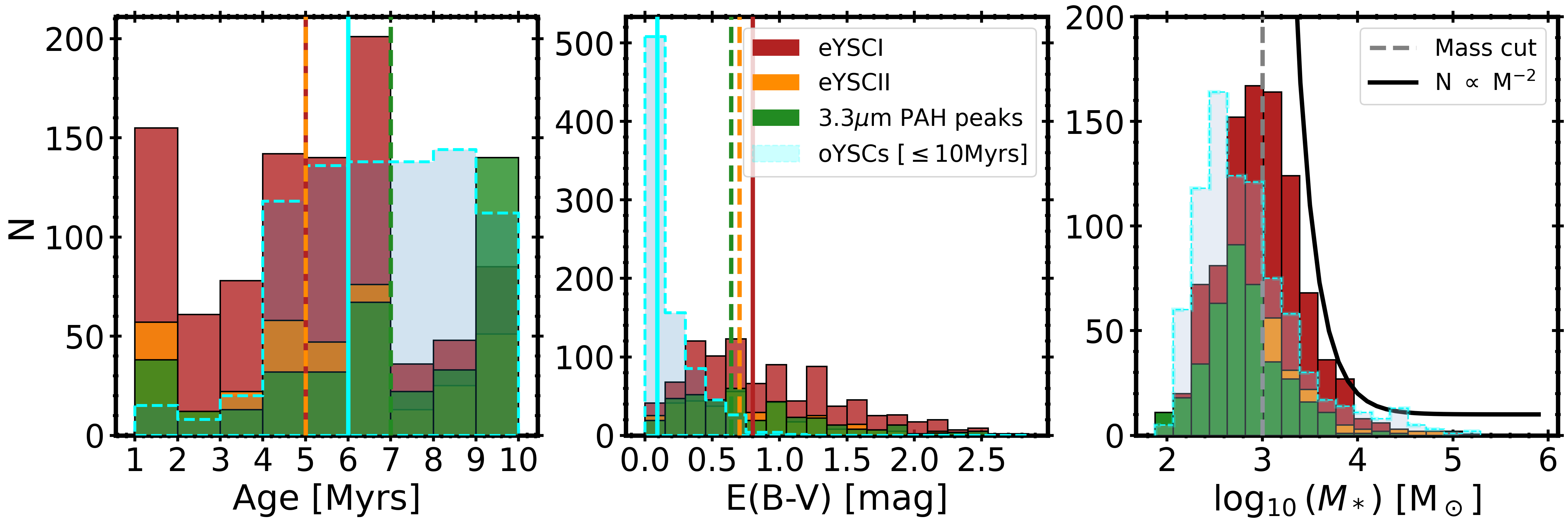}
    \caption{The stellar age (left), stellar attenuation (middle) and stellar mass distribution (right) for the different classes, \eYSCI\ (red), \eYSCII\ (orange), and \PAHclass\ (green) including the oYSCs ($\leq$ 10 Myr; cyan) derived with CIGALE. The vertical lines correspond to the median values of \eYSCI\ (red), \eYSCII\ (orange), \PAHclass\ (green), and the optical YSCS ($\leq$ 10 Myr; cyan). The mass distribution (right) also includes a grey dashed line corresponding to the mass cut at $10^3$ M$_\odot$ and the black curve to the 10-logarithm of a cluster mass function (CMF) with slope -2. For the stellar age (left panel) the \eYSCI\ median overlaps with the \eYSCII. } 
    \label{fig: distributions} 
\end{figure*}

In Figure \ref{fig: distributions}, we present the stellar age (left), color reddening distributions (middle), and mass (right) of each eYSC class and oYSCs, separately. We show median values of ages and E($B-V$) for each class as vertical lines.  We observe that the \eYSCI\ and \eYSCII\ are on average the youngest and most attenuated systems, followed by older and less attenuated oYSCs. The \PAHclass\ are at face value the oldest population.  We also notice that only a small fraction (5\%) of the oYSCs have ages below 3 Myr, while we find that this fraction increases to 30\% for \eYSCI, 25\% for \eYSCII\ and 16\% \PAHclass. On the right panel, we show the mass distributions for each eYSC class and the oYSCs. We include a power-law mass function (CMF) with slope -2 \citep{krumholz2019} to guide the reader to the expected shape of this distribution in the diagram. The oYSCs have a larger number of massive star clusters above $10^4$ \msun\ that are mainly associated with the central starburst region (see Section~\ref{sec: mass_dist_discussion}).  The turnover of the eYSC mass around $10^3$ \msun\ is due to incompleteness. The peak appears at lower masses in the oYSC population. Both populations follow the -$2$ power-law shape similarly reported by \citet{Levy2024} and  \citet{linden2024}. We find 48\% of \eYSCI,  39\% of \eYSCII\ and 25\% of the \PAHclass\ more massive than $10^3$ M$_\odot$, while only 30\% of the oYSCs have masses above this limit (illustrated by the gray vertical line in Figure \ref{fig: distributions}).

We note that stochastic IMF sampling is definitively a limitation for the recovered values of cluster physical properties. It is well known that the use of deterministic models (e.g., Yggdrasil, CIGALE, MAPPINGS) when fitting star cluster SED leads to relatively biased physical properties, especially as a function of cluster mass \citep[e.g.][]{cervino2004, maiz2010, fouesneau2010accounting, krumholz2015}. Here, a large fraction of clusters in all categories have masses below $10^3$ M$_\odot$, and thus we expect their ages to be unreliable due to stochastic IMF sampling. \cite{maiz2010} tested the effect of stochasticity using as baseline a SED with U--to--NIR broadband colors. They concluded that such an extended baseline is beneficial to recover a reliable cluster age, except in the red supergiant (RSG) phases around 10-30 Myr. \cite{fousneau2012} reached a similar conclusion using U--to--I bands, showing that RSG colors scatter the recovered ages of clusters, creating an artificial gap in the age distributions at 10-30 Myr and overpopulating the age range 6-10 Myr. With the addition of NIR broadband colors, \citet{fouesneau2010accounting} also the asymptotic giant branch (AGB) phases become critical for star cluster physical property recovery. These limitations are also inherited from the stellar evolutionary models that lack good prescriptions for these stellar phases. Related to the age range explored in this work, we report some important considerations. Since we limit the analysis to the cluster populations within the overlap area of HST and JWST FoVs, all cluster candidates in the emerging as well optical categories potentially have UV--to--5 $\mu$m detection. Adamo et al. and Pedrini et al. in prep suggest that some of the IR colors of the oYSCs as well as eYSCs suffer indeed from stochastic sampling effects. These works, however, highlight that the inclusion of 3 H recombination lines in the SED analysis (H$\alpha$, Pa$\alpha$, and Br$\alpha$) largely mitigate the age-extinction degeneracy that affects most of the broadband studies \citep[e.g][although the latter only included H$\alpha$ in their analysis]{calzetti2015, whitmore2020}. In particular, Pedrini et al. (in prep.) report a clear NIR excess in the eYSCs in the FEAST galaxies. This excess results in poor fits of the NIR continuum and overestimation of the eYSCs ages. Therefore, it is very likely that the true ages of the eYSCs are younger than reported by the SED analyses. The IR colors of eYSCs  clearly show Pa$\alpha$ excess with respect to the oYSC, confirming that the latter population is older compared to the recovered ages from the SED analysis. 

We speculate here that the difference in mass distributions between eYSCs and oYSCs might be due to uncertanties in the recovered ages but also to the dynamical properties of the environment probed by our observations and the limited area covered by the FoV. M83 is the only target among the FEAST galaxies featuring a strong bar (see Figure~\ref{fig: enviromental_map}). The JWST FoV encloses the bar, end of the bar, and the circum-nuclear starburst ring. These are the dustier regions in M83 and the areas where recent star formation is more enhanced. In Figure~\ref{fig: distributions_opticalfullfov} of Appendix~\ref{app:A} we show the ages, E(B$-$V), and masses of the oYSCs within the entire HST FoV. We see that the median age of the entire population becomes slightly younger (5 instead of 6 Myr), the median E(B$-$V) remains very low, and the mass distribution of the oYSCs peak at similar low masses as observed within the JWST FoV. However, we recover masses that are more similar to those observed in the eYSC populations. The latter suffers from significant incompleteness below $10^3$ \msun. It is possible that rapid disk rotation and limited FoV causes a selection bias against the optical YSC populations. The other possibility is that due to the dynamical timescales involved in the diverse galactic environments it might take longer for star clusters with masses above $10^3$ \msun\, to emerge from the parent cloud in M83. Finally, the difference in mass between eYSCs and YSCs might be due to mass loss on short timescales because a large fraction of eYSCs are probably expanding stellar associations. We discuss these possibilities in the next Section.

\section{Discussion} \label{sec: discussion}
In this study, we present the population of star clusters younger than 10 Myr in the star-forming galaxy M83. We extract the eYSC candidates in H-recombination emission line maps, \paalpha\ and \bralpha, and a \PAHlambda\ emission map tracing the ambient HII region and PDR, respectively. The YSCs are detected in the optical using the F555W and F547M HST filters. We perform multiwavelength photometry from the UV--to--5$\mu$m using the \texttt{FEAST-pipeline} and derive their physical properties using the SED-fitting code \texttt{CIGALE}. The result is 1307 eYSCs, 389 \PAHclass\ and 829 oYSCs residing within the JWST NIRCam FoV.

\subsection{The \PAHclass}

In Section~\ref{sec: results}, we presented  the colors and physical properties derived for the \PAHclass, identified as compact emission sources in \PAHlambda\, emission but not significantly in H recombination lines. The colors of the \PAHclass\ (Figure~\ref{fig: colorcolorplot_double}) can be explained by  strong emission in the  PAH band,  while, compared to the eYSCs,  they have on average the lowest emission in Pa$\alpha$.  Their colors are also different from the eYSC-BB suggesting that these are not (only) deeply embedded clusters. Focusing on CIGALE SED outputs (Figure~\ref{fig: distributions}), we see that in general this class is associated to older ages, similar attenuation but lower masses than eYSCI. The lack of a compact/strong HII region associated with these systems might lead deterministic models to age-date these systems as older on average. We speculate here that these systems might be those clusters that do not host stars massive enough (detection limit correspond to a 15 \msun\ or O9.5V spectral type) to power a detectable HII region. PAH molecules are excited by non-ionizing UV radiation and for a population of B stars could still result in detectable PAH emission. Both explanations, older systems or young but not hosting massive stars, could explain this population. Due to their uncertain nature, we have reported their detection and derived physical properties, but we will not include them among the eYSCs in the following discussion.

\subsection{Are all eYSCs gravitationally bound?} \label{sec: boundvsunbound}

A stellar cluster, as mentioned in Introduction, can be gravitationally bound or unbound. For example, the YSCs classified according to the LEGUS scheme \citep{adamo2017} as classes 1 and 2 have higher probability to be gravitationally bound, while class 3 are more likely stellar associations. This morphological classification has been based on the measurements of the dynamical age, defined as the ratio between the cluster age and its crossing time \citep{Gieles2011, Ryon_2017, brown2021}. However, this method does not work on star clusters that are too young, since the crossing time becomes comparable to the stellar age  \citep{Gieles2011}. Star formation is clustered in nature, which challenges our understanding of the formation of gravitational bound star clusters with respect to stellar associations. Initial numerical simulations suggested that the formation of a stellar association was the result of an expanded bound cluster, unbounded by rapid gas dispersion \citep[e.g][]{kroupa2001, goodwin2006gas}. The implications of these works are that the majority of star formation would take place in bound clusters. However, more realistic initial conditions for a GMC collapse can, even at the start, produce stellar associations \citep[e.g][]{parker2014dynamical, Dobbs2022, grudic2022}. Some recent observational studies \citep[e.g][]{kounkel2018apogee, ward2020not} also suggest that expanding associations are initially more compact but still unbound. This is further supported by the fact that a large majority of young systems, i.e. embedded clusters, are actively expanding \citep{Kuhn_2019}. 

In this work, we can use eYSCs and oYSCs relative numbers as an indication of what fraction of eYSCs we should expect to be potentially bound. While we cannot yet establish the eYSC boundness using for example their dynamical age, we can compare total numbers of eYSCs and oYSCs to derive a first order estimate of the fraction of bound clusters within the eYSCs. In this exercise we assume that the galaxy star formation rate (SFR) is constant within 10 Myr, that eYSCs are precursors of oYSCs and associations dissolve within 10 Myr, which is reasonable since we look at compact systems within a few parsec scales. Among the oYSCs younger than 10 Myr we have classified $\sim$36\% (299  oYSCs) as likely bound (class 1 and 2), while the remaining 64 \% are short-lived associations (530 systems as class 3). If we compare the number of bound oYSCs with the total number of eYSCs (eYSCI$+$eYSCII) we find that: N(YSCs (class 1 and 2))/N(\eYSCI+\eYSCII) $\sim$ 23\%. In other words, only between 20--30\% of the eYSCs could potentially be bound while the majority, 70--80\%,  are likely associations.  This result agrees with observations in the Milky Way \citep{Lada} where most of the stellar associations dissolve within 10 Myr as well as  numerical works such as \cite{Farias2024}, who found that most of the star clusters formed in the STARFORGE simulations are unbound.

\subsection{Physical properties of eYSCs and YSCs as a function of galactic environments} \label{sec: mass_dist_discussion}

Previous studies of the YSC \citep{bastian2012, adamo2015} and GMC \citep{freeman2017varying} populations of M83 find that their mass distributions can be described by a power law with slope -2, while the upper-mass end is consistent with an exponential cut-off mass that changes in different galactic environments. These studies pointed out that the nuclear starburst and the end of the bar regions appear to be more efficient in forming more massive clusters in agreement with analytical and numerical model predictions \citep{reinacampos2017, Ahmad2023}. 

The eYSC population detected in this work is the connecting step between oYSCs and GMCs. We therefore investigate here whether the mass distributions of the eYSCs show consistent variations. First we divide the FoV in radial annuli as a function of distance from the center in an informed way. Each annulus contains a specific environment (see Figure \ref{fig: enviromental_map}): the center within 0.34~kpc, the bar between $0.34$ and $R_g$~$<2.1$ kpc, the end of the bar connecting with the spiral arm at $2.1$~$\leq$~$R_g$~$<3.3$ kpc, and the outer regions. Since we are focusing on stellar populations formed within 10~Myr, we expect that each annulus contains systems formed under the same physical conditions. The number of eYSCs and oYSCs in each region are reported in Table~\ref{tab: timescales}. 

\begin{figure}
    \centering
    \includegraphics[width=1\linewidth]{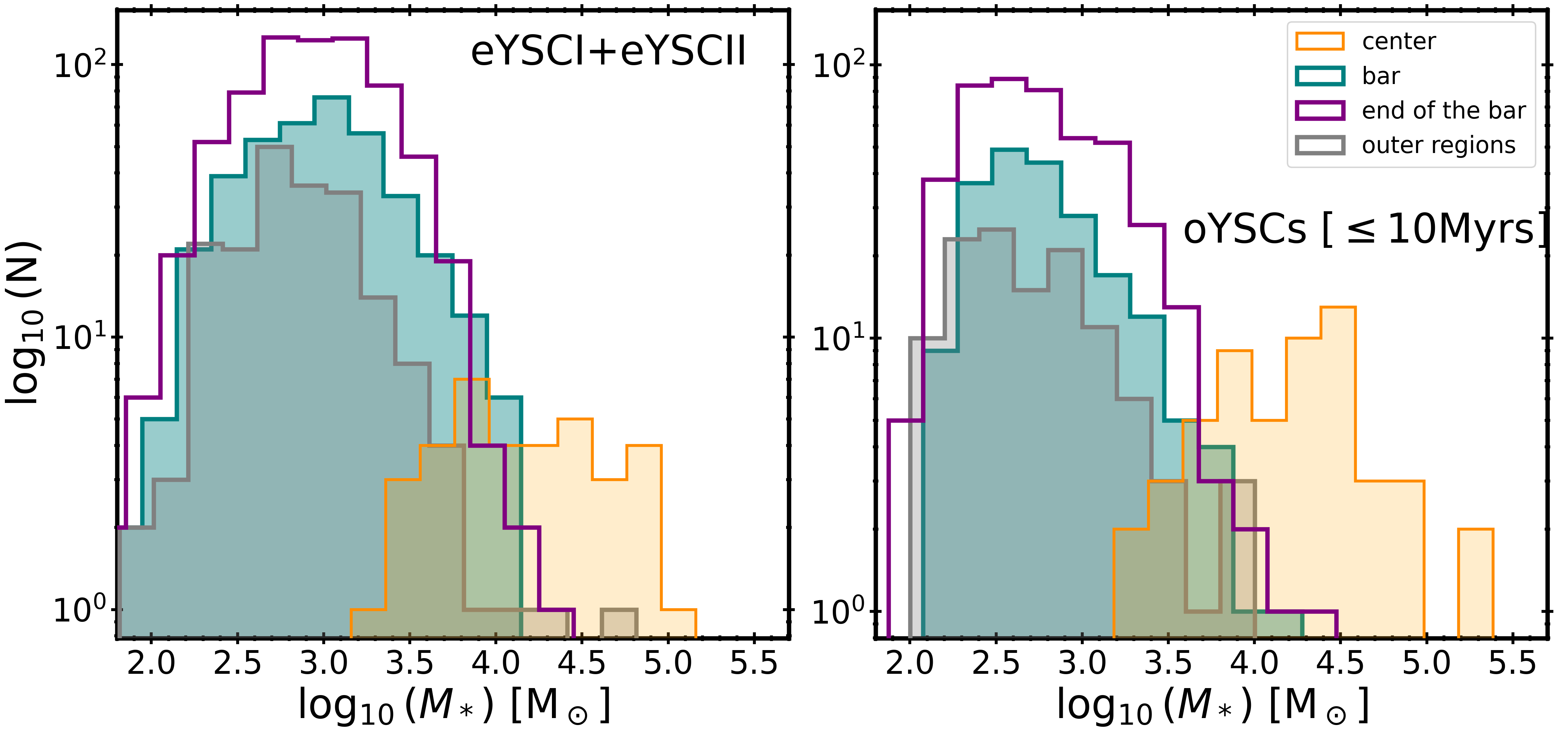}
    \includegraphics[width=1\linewidth]{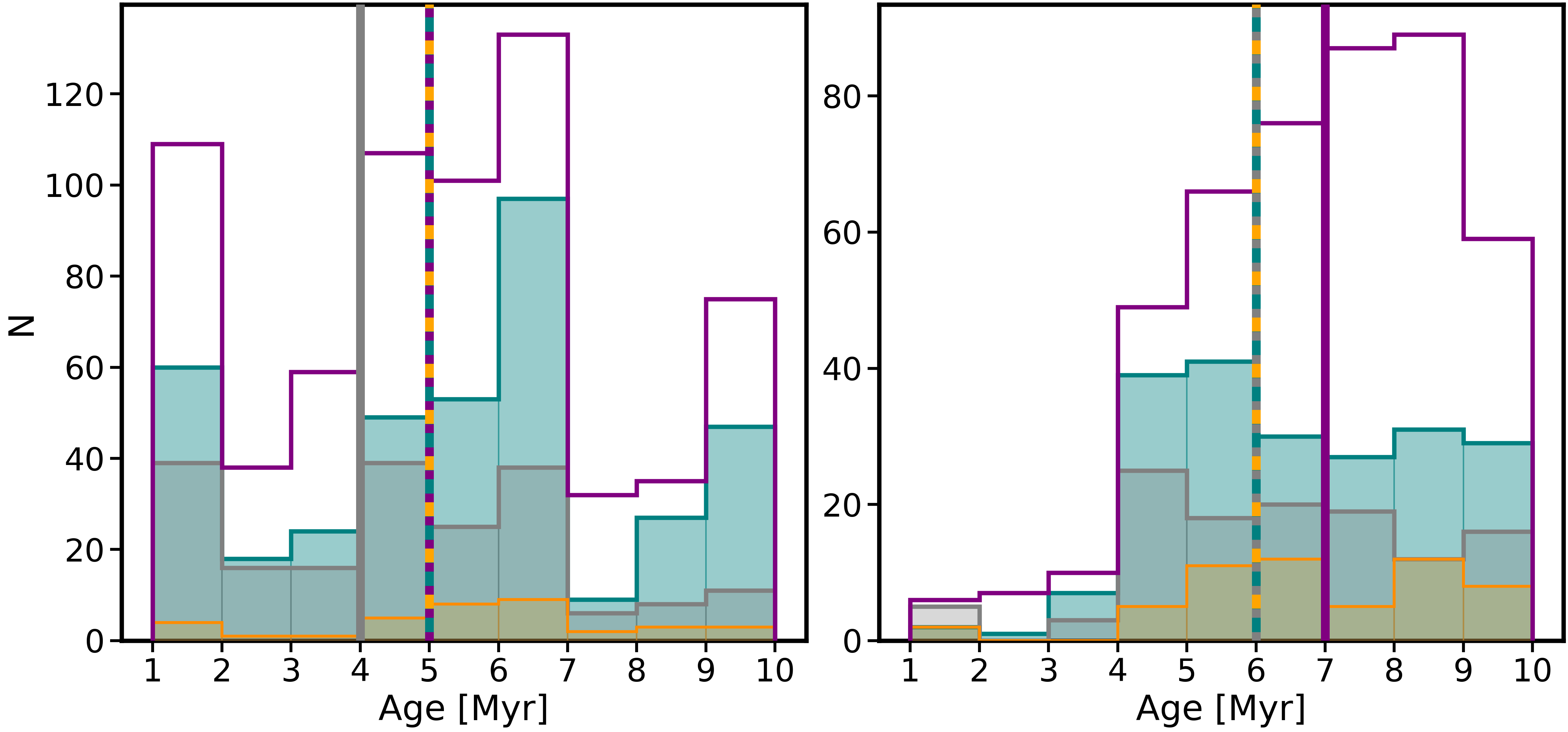}
    \caption{The mass (top) and age (bottom) distributions of the combined \eYSCI\ and \eYSCII\ sample (left) and oYSCs  (right) as a function of different galactic environments selected to encompass different dynamical regions of the galaxy: Center (yellow), Bar (teal), End of the bar (purple), Outer regions (grey). The turnover in the mass distributions is due to incompleteness. The vertical lines in the bottom plot show median ages in each region.}
    \label{fig: informed_env}
\end{figure} 

\begin{figure}
    \centering
    \includegraphics[width=1\linewidth]{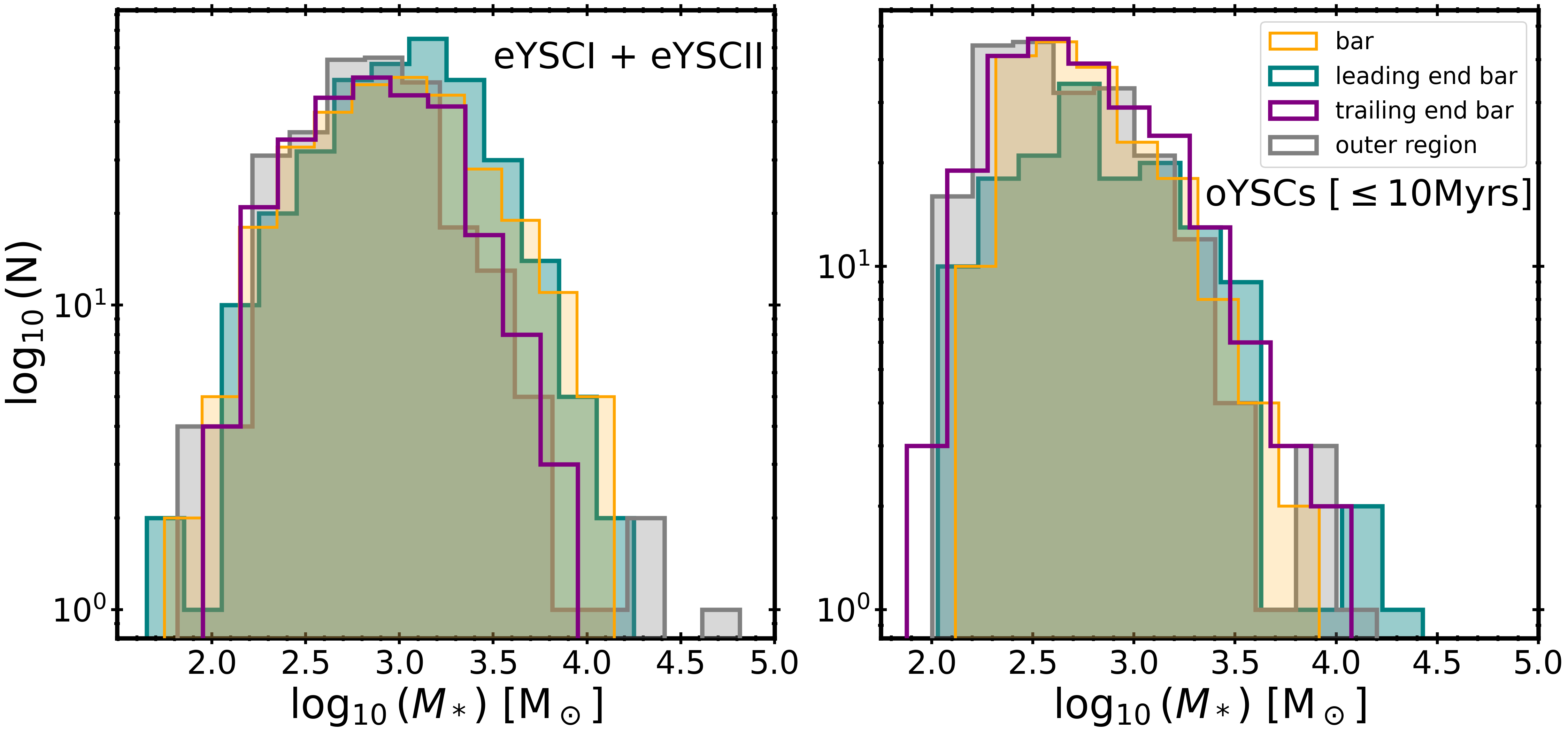}
    \includegraphics[width=1\linewidth]{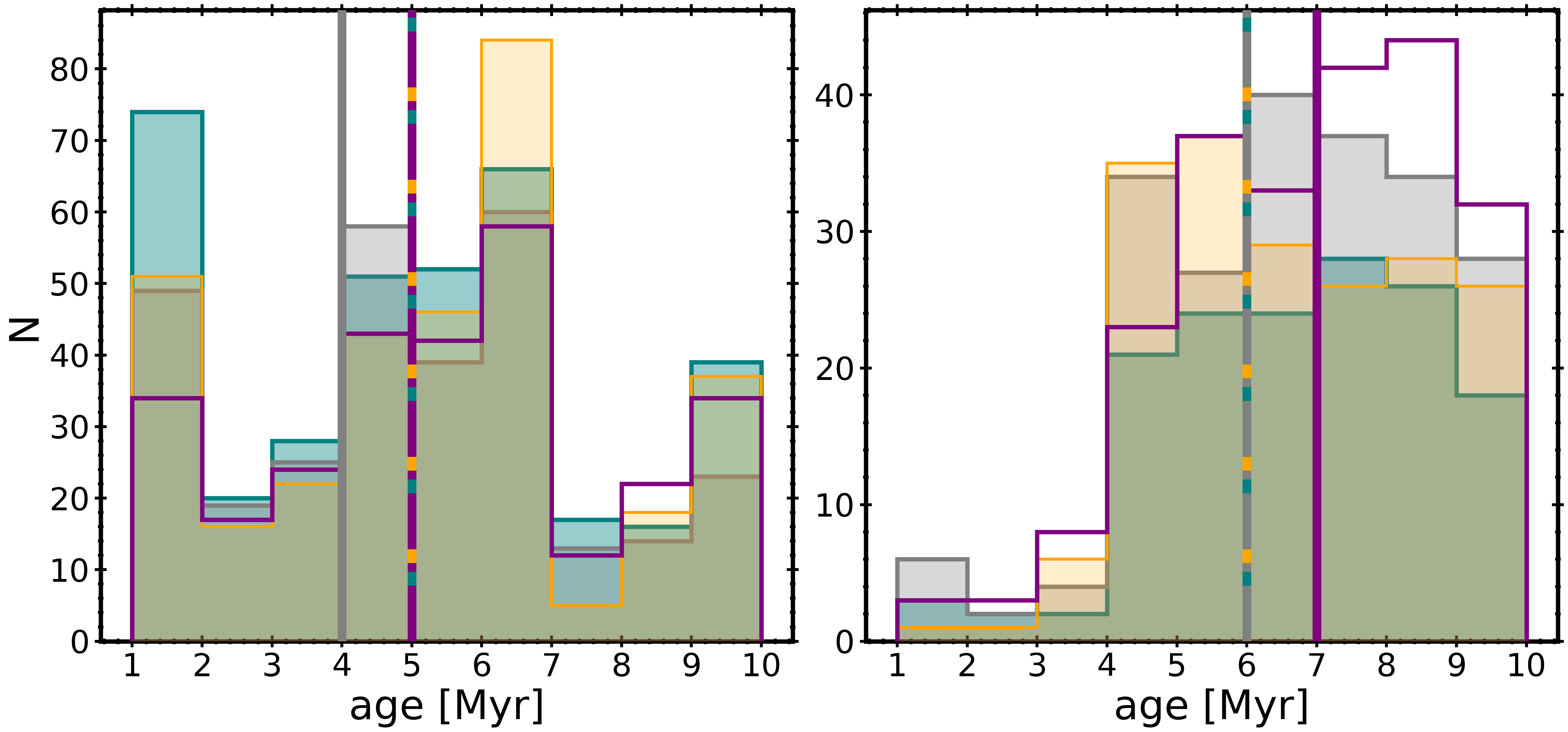}
    \caption{The mass (top) and age (bottom) distributions of the combined \eYSCI\ and \eYSCII\ sample (left) and oYSCs  (right) in bins containing the same number of eYSCs and oYSCs combined. The central starburst region has been removed. This alternative division is less sensitive to changes in SFR due to different dynamical properties. The turnover in the mass distributions is due to incompleteness. The vertical lines in the bottom plot show median ages in each region.  }
        \label{fig: equnum_disk}
\end{figure} 

In Figure~\ref{fig: informed_env} we show the age (bottom) and mass (top) distributions of the combined eYSCs (left) and of the oYSCs (right) for each informed environment. Overall the median ages of the eYSCs and YSCs in the different environments coincide with those seen in the respective populations in Figure~\ref{fig: distributions}, with the exception of eYSCs in the outer regions appearing slightly younger and the oYSCs at the end of the bar annulus being slightly older than average. In the top plots, the flattening and turnover of the distributions show at which mass incompleteness become severe. The centre, bar and end of the bar are the most affected, while the outer regions have lower mass incompleteness. The eYSC and oYSC mass distributions in the centre stand out with respect to the rest of the disk. While incompleteness is severe, we also see that this region is the one where the most massive clusters are currently forming, confirming previous results in this galaxy \citep[e.g.][]{harris2001} and in nuclear starburst ring in general \citep[e.g.][]{vanderlaan2015}.  The bar and end of the bar region contains the largest number of eYSCs and oYSCs and samples clusters up to a few $10^4$~\msun. The outer regions contain a modest number and have the least massive clusters. These trends are very similar to what reported by \cite{Ahmad2023}, who simulated star cluster formation in clouds extracted in diverse galactic environments. They find that clusters forming in molecular clouds affected by bar potential are able to sample more massive star clusters than in the spiral arm and inter--arm region.

We try next to assess whether the dynamical features in the disk of the galaxy (e.g. bar, spiral arm, inter-arm) organize cluster formation, in the same way they organize star formation. We divide the galaxy disk area outside of the centre in annuli containing the same number of eYSCs$+$oYSCs (see blue circles in Figure~\ref{fig: enviromental_map}). Assuming that cluster formation is a constant fraction of the SFR in the last 10 Myr, by dividing the regions based on the same number of clusters should overcome SFR variations. Overall, this division produces annuli covering the bar and the outer region in a similar fashion as before, but it divides the end of the bar region into two: We refer to these new annuli as the leading-- (closer to the bar) and trailing -- (closer to the spiral arm) end of the bar.  In Figure~\ref{fig: equnum_disk}, we plot the mass and age distributions in these alternative annuli. The mass distributions of the eYSCs in the bar and leading--end of the bar annuli have more massive clusters than the trailing and outer region annuli (with some exception for the latter). These differences are not as clear in the mass distributions of the  YSCs. Interestingly, when looking at the age distributions in Figure~\ref{fig: informed_env}, we see that the median age distribution of the eYSCs do not change with respect to the previous division. On the other hand, the division of the bar highlights that the average older population seen in the oYSCs (bottom right panel of Figure~\ref{fig: informed_env}) is preferentially associated to the trailing side of the bar connecting with the spiral arm, likely due to strong dynamical drifting associated with that region. 

We conclude here that while the centre stands out for its ability to form very massive star clusters, the differences seen in the eYSC and YSC mass distributions in the disks are likely driven by the increase in SFR. The latter results in a sampling of the mass function to higher masses. This result is in agreement with recent analyses of molecular clouds in local spiral galaxies \citep[e.g.][]{sun2020, querejeta2024, faustino2025} that find noticeable differences between the molecular cloud properties in the centre of the galaxies while these differences become smaller in the disk, i.e. the spiral arms appear to organize star formation but not enhance it. We also notice that the limited coverages, reduces the sampling of the disk environment, especially the arm, inter-arm regions and therefore, our conclusions might be biased by limited statistics. In a follow-up work, we will provide an in-depth analysis of the eYSCs and YSCs population across the FEAST galaxies to test environmental dependencies on the shape of the mass function.

\subsection{The Emerging sequence of star clusters} \label{sec: sequence}

One of the main goals of this study was to estimate the emerging timescale: that is the time it takes for an eYSC to disperse its natal cloud. These timescales are fundamental to set the integrated star formation efficiency in the GMCs where star clusters are forming as well as the timescales for stellar feedback to disrupt the natal GMC.

In Section~\ref{sec: results} we found  that \eYSCI\ and \eYSCII\ are younger and more attenuated than the oYSCs.  In Figure \ref{fig: colorcolorplot_double} (upper right) we see evidence that the eYSCs follow an evolutionary track with the \eYSCII\ located at bluer IR colors and, therefore, being the least embedded, followed by the 3.3 $\mu$m PAH peaks and lastly, the eYSCI at redder color, being the most embedded star clusters detected in this study.  

Using a broadband color selection aimed to find the most embedded star clusters not necessarily associated to Pa$\alpha$ emission, resulted in the extraction of only 70 sources not yet classified as eYSCs (about 5\%). This small fraction, similarly to what is found in NGC628 (Adamo et al. in prep.), would suggest that the deeply embedded phase, prior to the massive stars reaching the main-sequence and starting to power a detectable HII regions, is significantly shorter when compared to the phase dominated by ionised gas emission. The \PAHclass\ could potentially contain deeply embedded clusters or clusters that did not form massive stars. However, we see that only 15 \% of this class has colors compatible with this selection, reinforcing the conclusion that the deeply embedded phase is short when compared to the other phases. This evidence is in agreement with high-spatial resolution studies of embedded cores conducted in the Milky Way \citep{evans2009, DC2013}, where the accreting protostellar phase (i.e. corresponding to Class 0/I YSOs) appears to be very short ($<$1Myr).

The evolutionary sequence identified here is mainly based on the PAH emission feature prominence between classes, where the \eYSCI\ have a compact and stronger PAH emission (brighter F335M magnitude) than \eYSCII. This difference would imply an evolution in the morphology of the PDR associated to the \PAHlambda\ emission. \cite{pedrini2024feast} find evidence that supports this hypothesis by reporting differences between the PDR morphology for classes of eYSCs and, especially, a decrease in the \PAHlambda\ emission feature with cluster age. In conclusion, the physical properties of \eYSCI, \eYSCII, and oYSCs argue for a complete sampling of the evolutionary sequence from deeply embedded to exposed.  As discussed in Adamo et al (in prep.), studies based on JWST broadband color selections and the excess of \PAHlambda\ emission \citep{rodriguez2023, rodriguez2024, Whitmore_2023, Linden_2023, linden2024, Levy2024} are sensitive to the embedded phase, e.g. traced here by the eYSCI class, but miss the eYSCII class. Pre-JWST studies based on the intensity of H recombination lines \citep[e.g. H$\alpha$, Pa$\beta$]{hollyhead2015studying, Hannon2019, hannon22, Messa2021, mcquaid2024, deshmukh2024} miss a significant fraction of embedded objects (\eYSCI) due to progressively higher extinction, but also to the lower resolution
of previous NIR studies, which are now revealed by the more sensitive JWST data as well as sampling up to 5~$\mu$m. Incomplete sampling of either these evolutionary phases leads to shorter timescales. We, therefore, use the emerging YSCs in M83 to derive a more reliable timescale than previous studies.

\subsection{Cluster physical properties along the emerging sequence} \label{sec: trends}

\begin{figure*}
   \hspace*{-1cm}%
    \centering
    \includegraphics[width=\linewidth]{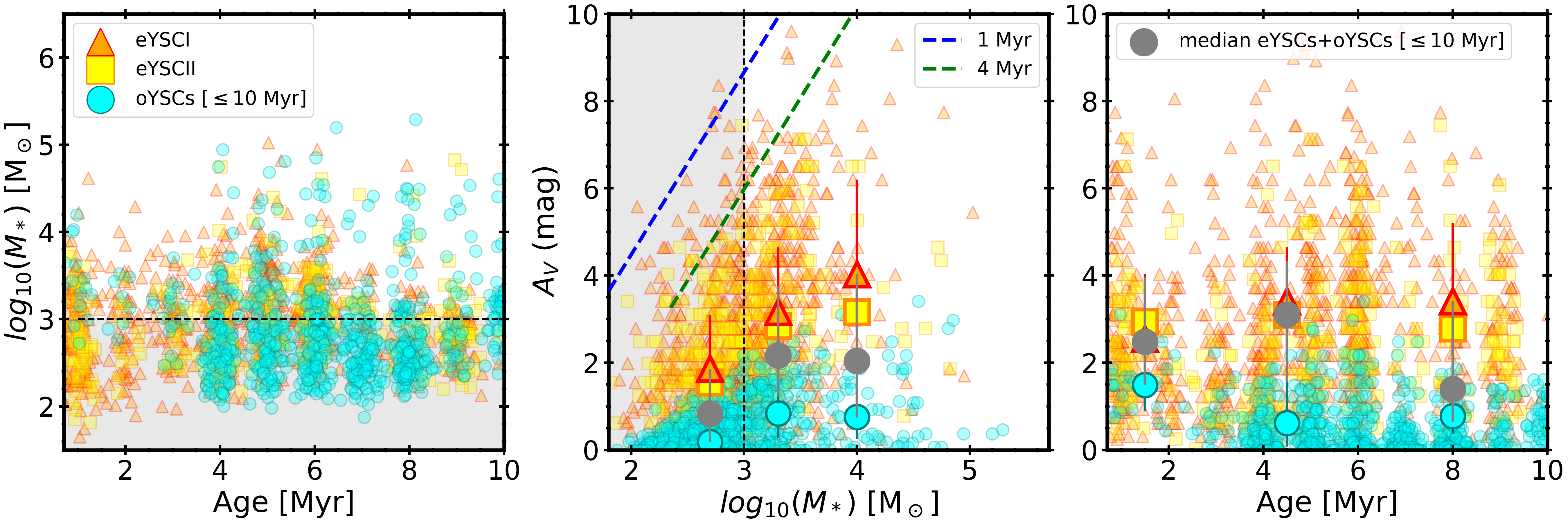}
    \caption{Physical properties of the combined cluster sample younger than 10 Myr. On the left, we show the age-mass distributions. The total visual extinction ($A_V$) is plotted against the stellar mass (middle) and age (right). We can observe the \eYSCI\ (orange triangles), \eYSCII\ (yellow squares), YSCs ($\leq$ 10 Myr; blue circles). Large symbols show the median $A_V$ for the \eYSCI, \eYSCII, and YSCs ($\leq$ 10 Myr) separately and combined (grey dots). In the middle plot, we use $10^3$$-$5$\times10^3$ and $>$ 5$\cdot10^3$ M$_\odot$ mass bins to estimate median  $A_V$. The dashed lines corresponds to the detection limits for the maximum $A_V$ as a function of stellar mass for star cluster with age 1 Myr (blue) and 4 Myr (green) estimated using Yggdrasil models. In the right plot, we estimate the median $A_V$ in age bins 1$-$3, 3$-$6, 6$-$10 Myr after the mass cut, to mitigate the effect of stochastic sampling in the age derivation. The gray areas highlight star clusters with mass below 1000 \msun.}
    \label{fig: ext_trends}
\end{figure*} 

Combining HST narrow band Pa$\beta$ with UV--optical broadband imaging, \cite{Messa2021}, \cite{mcquaid2024}, and \cite{deshmukh2024} reported a weak anti-correlation between age and extinction, with a significant fraction of clusters associated with Pa$\beta$ emission having only moderate extinction already from early ages. In NGC4449, \cite{mcquaid2024} found that more massive clusters are associated with lower attenuation, suggesting that the clearing timescales could be shorter for more massive cluster mass. Following these previous analyses, we plot the age and mass distributions of eYSCs and YSCs (left), the total visual extinction against the cluster mass (centre), and the stellar age (right) in Figure \ref{fig: ext_trends}. The left plot shows that similar trends as seen in Figure~\ref{fig: distributions}, with eYSCs dominating the youngest age bins and the opposite is true for the oYSCs. The gray area show highlight clusters with masses below 1000 \msun\, where stochastic IMF effects will dominate. In the central and right panel, we plot $A_V$, estimated using the stellar extinction, E(B$-$V), outputted by CIGALE, and the gradient of the extinction curve at visible wavelength, $R_V=3.1$ \citep{ISMbook}. In the central panel, we include median and quartiles of $A_V$ as a function of mass ($\leq10^3$, $10^3$--$5\times10^3$, $>$$5\times10^3$ M$_\odot$) bins for each eYSC class and oYSCs as well their combined values (grey dots).  We also show, as a guidance, the detection limits for the maximum $A_V$ as a function of cluster mass, assuming two representative age values estimated using \paalpha\ detection limits and Yggdrasil's models, to indicate how incompleteness affects the distributions. In all classes, low-mass clusters will be affected more by incompleteness at increasing values of $A_V$ than massive ones. Although we include one bin below 1000 \msun (gray region), it is important to remember that that bin is the one most affected by incompleteness and uncertanties due to stochastic IMF sampling. By focusing on median trends of $A_V$ in the 2 mass bins above 1000 \msun, we do not see a clear correlation between mass and extinction in oYSCs. The eYSCs show a large scatter in $A_V$ as a function of mass and a tentative positive correlation, i.e. that more massive clusters might be more attenuated. However, when combining the NIR and optical populations (grey dots) and taking into account also the incompletness we conclude that there is no evidence of a strong trend. We do not find a clear signal that clearing timescales might depend on cluster mass. This result might be driven by the uncertainties and biases in the recovered cluster physical properties or might be due to the assumption of a simple screen attenuation which might not truly reflect the true dust geometry. 

We finally focus on the $A_V$ as a function of age in the right panel of Figure~\ref{fig: ext_trends}. We include median estimates of $A_V$ as a function of 3 age (1--3, 3--6, 6--10 Myr) bins using only clusters more massive than $10^3$ \msun\ to mitigate strong deviations by stochastic effects. We do not observe a clear trend for the eYSCs: the attenuation with age is consistent within errors. On the other hand, the optical counterpart of the YSCs have moderate attenuation during the first few Myr which declines to almost zero. When we combine the two populations, we see a clear declining trend. At face-value, the different behavior between the two populations could indicate orientation and projection effects at play. The first age bin is dominated by eYSCs and only a smaller fraction of YSCs. We see here that the clusters are associated on average (grey dots) with E(B$-$V)$\sim$1 mag. Only after 4 Myr, the overall attenuation decline goes to E(B$-$V)$\sim$0.3 mag in the last age bin. We conclude that while the general trend suggests clearing timescales of 5-6 Myr, about 20\% of attenuated eYSCs might still be present after 6 Myr. In Pedrini et al. (in prep) and Linden et al. (in prep.) we find evidence that these older eYSCs associated with high E(B$-$V) might not be correctly fitted by CIGALE. We notice that these eYSCs still have strong Pa$\alpha$ and H$\alpha$ associated to them and EWs estimates would suggest they are younger than 5 Myr. Taking into account these uncertainties, the trend presented above becomes stronger: newly formed clusters go through an emerging phase initially associated with higher attenuation along the line of sight, which declines at older ages. The large scatter might be due to degeneracies between orientation effects as well as to uncertanties to where exactly attenuation is taking place along the line of sight. A similar trend in attenuation has also been reported for larger star-forming regions in NGC628 by \cite{pedrini2024feast}, where the classification was based on the PAH morphology.  

\subsection{The emerging timescales}
Another way to derive emerging timescales is not to rely on inferred physical properties for the eYSCs but to look at the number fractions of star clusters in different emerging stages assuming a timescale of 10~Myr \citep[e.g.][]{Whitmore_2023}.  
For this exercise, we use the number of \eYSCI, \eYSCII, and  oYSCs with solid detections and good $\chi_{\text{red}}^2$  (see table \ref{tab: final_cat_1}) and we limit the oYSC population within the JWST NIRCam FoV. The timescale is given as: 
\begin{equation}\label{eq: timescale}
 t=\frac{\text{N(eYSCI)+N(eYSCII)}}{\text{N(eYSCI)+N(eYSCII)+N(oYSCs)}}\cdot \text{\emph{age}}   
\end{equation}
where $N(X)$ corresponds to the number of $X$ type clusters, \eYSCI, \eYSCII, or the  oYSCs, and the age parameter is 10~Myr. This method makes the implicit assumption that the star formation rate is constant over the period considered. The total emerging timescale corresponds to the time period for the \eYSCI\ and \eYSCII\ to become YSCs. We also estimate in a similar way the length of the \eYSCI\ phase by dividing only the number of objects in this phase to the total number of clusters.  

The recovered timescales are reported in Table~\ref{tab: timescales} for the galaxy region sampled by JWST and the different galactic environments introduced in Section~\ref{sec: mass_dist_discussion}. The errors are estimated using Poisson statistics and error propagation for the number of clusters. Unless specified, we do not apply any mass selection above 1000~\msun. By selecting only eYSCs and oYSCs more massive than 1000 \msun\, we recovered timescales and trends consistent to those reported in Table~\ref{tab: timescales}. 
The emerging timescale for the covered portion of the galaxy is $6.1$~$\pm$ $0.1$~Myr. This is on average the time that takes a star cluster to go from embedded to fully exposed in M83. These timescales are similar to the ones derived above by comparing $A_V$ versus age. They are longer than reported  by \cite{hollyhead2015studying} in M83, where using H$\alpha$ morphology combined with cluster age they find that the clusters have dispersed their natal cloud by $4-5$~Myr. 
The slightly longer timescales can be understood here by the more complete sampling of the emerging phase accessible with JWST.

The first phase, from embedded to not being associated to a compact PDR, e.g. eYSCI phase, lasts about $4.4$~$\pm$~$0.1$ Myr. This timescale is comparable to those reported in the PHANGS galaxies \citep{Whitmore_2023, rodriguez2023, rodriguez2024} and the GOALS starbursts \citep{Linden_2023, linden2024} where the \PAHlambda\ emission has been used as a tracer.

Once a compact PDR is disrupted, it takes only about 2~Myr for an average cluster to not be associated with a compact HII region, e.g., eYSCII phase. The short time spent in the eYSCII phase can explain the similar age distributions for the \eYSCI\ and \eYSCII\ (see Figure~\ref{fig: distributions}, left).  These short timescales are similar to those reported in M83 by \cite{hollyhead2015studying}. The latter  find that a cluster takes about 2~Myr to go from partially embedded to exposed phases. \cite{deshmukh2024} recovered an emerging timescale of 2-3~Myr for Pa$\alpha$ identified clusters, similar to what we find for eYSCII. 

Studies where molecular gas tracers are combined with mid--IR and UV--optical wavelengths recover consistent timescales as we find here \citep[e.g.][]{corbelli2017, kim2023, Grasha2018, Grasha2019}. In particular, \cite{Sun_2024} using high-spatial resolution observations with ALMA, HST, and JWST in the centre of NGC3351, report a starless phase lasting 1-2~Myr and a total time for the emergence sequence of 4-6~Myr. 

In Table~\ref{tab: timescales} we also report the timescales dividing the sample in mass bins. We overcome incompleteness and stochastic IMF sampling issues by looking at clusters between 1 and 5$\times10^3$~\msun\ which provides characteristic timescales for the average star cluster of a few $>10^3$ \msun and at clusters more massive than 5$\times10^3$~\msun\ to provide representative  timescales for  $\sim10^4$~\msun\ clusters. We find that in general, for low mass clusters the complete emergence timescales goes up to 7~Myr, while for the massive clusters it goes down to 5~Myr. We argue that this result is not in tension with the lack of a strong trend between $A_V$ vs cluster mass in Figure~\ref{fig: ext_trends} because orientation effects and the simplistic assumption of a dust screen geometry might affect any correlation there. 

\begin{deluxetable*}{lccc}
\tabletypesize{\scriptsize}
\tablewidth{0pt} 
\tablecaption{The emerging and \eYSCI\ timescale for the different galactic environments presented in Figure \ref{fig: enviromental_map}, together with the galactocentric distances enclosing the region and the number of \eYSCI, \eYSCI, and oYSCs in each environment.  
\label{tab: timescales}}
\tablehead{
\colhead{Sample} & \colhead{eYSCI $\rightarrow$ oYSC}& \colhead{eYSCI $\rightarrow$ eYSCII} & \colhead{eYSC I+II, oYSC }\\
\colhead{M [\msun] or $R_g$ [kpc]} & \colhead{[Myr]} & \colhead{[Myr]} & \colhead{\#}} 
\startdata 
Whole galaxy & $6.1\pm0.1$ & $4.4\pm0.1$ & 946+361, 829 \\
Low mass (1e3 $<M\leq$ 5e3) & $7.4\pm0.2$ & $5.7\pm0.2$ & 394+121, 181 \\
High mass( $M>$5e3) & $5.3\pm 0.4$ & $4.0\pm 0.4$ & 56+19, 66\\
\hline
\hline
\multicolumn{4}{@{}c@{}}{Informed environmental division}\\
\hline
Center [$R_g$ $<$ 0.34] & $4.0\pm0.5$ & $2.4\pm0.5$ &  22+14, 55 \\
Bar [0.34 $>$ $R_g$ $\leq$ 2.1]& $6.5\pm0.2$ & $4.3\pm0.2$ &  256+128, 207\\
End of the bar [2.1 $<$ $R_g$ $\leq$ 3.3]& $6.1\pm0.1$ & $4.6\pm0.1$ & 521+168, 449 \\
Outer regions [$R_g$ $>$ 3.3] & $6.3\pm0.3$ & $4.7\pm0.3$ & 147+51, 118 \\
\hline
\hline
\multicolumn{4}{@{}c@{}}{Regions containing same number of eYSCs+oYSCs}\\
\hline
Bar [0.34 $<$ $R_g$ $\leq$ 1.9]& 6.3 $\pm$ 0.2 &  4.2 $\pm$ 0.2  & 214+108 , 189 \\
Leading end of the bar  [1.9 $<$ $R_g$ $\leq$ 2.5]  &7.1 $\pm$ 0.2 &  5.2 $\pm$ 0.2 & 276+96, 148\\
Trailing end of the bar  [2.6 $<$ $R_g$ $\leq$ 2.9] & 5.6 $\pm$ 0.2 &  4.2 $\pm$ 0.2 & 212+74 , 225\\
Outer region [$R_g$ $>$ 3.0] & 5.9 $\pm$  0.2 & 4.5 $\pm$ 0.2 & 231+69, 212 \\
\enddata
\end{deluxetable*}

When focusing on different galactic environments sampled in the FoV, we notice that the overall emerging timescales and the initial phase, traced by eYSCI, are comparable except in the center of the galaxy (see Table~\ref{tab: timescales}). When using binning of equal numbers of combined eYSCs$+$YSC, we see that the emergence timescales go up to 7 Myr in the leading side of the end of the bar region. This region is dominated by eYSCs, while the trailing portion of the end of the bar, the one connected to the spiral arm, show the opposite trend. As already mentioned above strong drifting coinciding with this region probably drives the differences in the timescales, the latter reflects dynamical environmental effects more than average timescales. 

The center of M83 stands out with respect to the rest of the galaxy disk for its elevated star formation activity and the presence of a large fraction of the massive star clusters in the sample. The star formation rate in the center is significantly higher than in the disk \citep[e.g.][]{adamo2015}.  \cite{callanan2021centres}, combining detection of star clusters in different tracers, propose that the star formation rate in the center has a variability timescale shorter than 10 Myr due to the dynamics of the bar supplying gas toward the nuclear ring. They estimate that the last peak event occurred 5-7 Myr ago, producing the massive YSCs we observe today, while the region is currently in a lower star formation mode. We indeed find that in the central region the number of oYSCs is larger than the detected eYSCs. However, we notice that this observational trend can also be explained if the emerging phase of these massive eYSCs is faster, leading to a larger number of oYSCs vs. eYSCs. 
We will investigate the mass dependence of the star cluster emergence phase with stronger statistics by combining the YSC and eYSC populations for all the FEAST galaxies in an upcoming work (Pedrini et al. in prep.).

In the next Section we will discuss the emergence phase by comparing these results to observational and numerical results in the literature.

\subsection{What stellar feedback drives the emerging phase of star clusters?}

Our analysis of the M83 cluster population suggests average emergence timescales of 6-7~Myr for typical cluster masses of a few 1000~\msun. We see initial evidence that the emergence timescales are shorter ($\sim5$Myr) for clusters above 5000~\msun. This average timescales do not include the deeply embedded phases of clusters where proto--stars are still accreting. However, as seen in the Milky Way infrared dark clouds, this phase is very short \citep[$<1$Myr, e.g.][]{evans2009}. Simulations of molecular clouds might help here to shed light on the dominant mechanism leading to cluster formation and feedback. Unfortunately, we do not have direct observational information of the physical properties of the molecular clouds that leads to the formation of the star clusters in M83, but we refer to the cluster mass end products to compare observations with simulations. The initial conditions of the molecular cloud, especially the density, have large impact on the resulting cluster mass and net star formation efficiency. The latter quantity and the type of stellar feedback are tightly intertwined and will determine the emergence timescale (cloud disruption in simulations) and whether clusters form bound as an end product \citep[e.g.][]{kim2018,  grudic2021, fukushima2022}. In a recent work by \citet{polak2024}, the authors find that denser/more massive clouds lead to the formation of more massive star clusters. Star formation efficiency is higher since radiative feedback  and stellar winds remain inefficient to prevent the collapse. Their formation is also faster and their emergence timescales shorter. Similar results are obtained by \citet{kim2018, fukushima2022, Menon2023} among many others.

It is also important to compare the role that pre-SN feedback plays in shaping the cluster properties and lead to the disruption of the natal molecular cloud.

Assuming a very short, deeply embedded phase, emergence timescales of 5 to 7 Myr might imply that SNe feedback is necessary to clear the leftover material from the natal cloud, since SNe explosions are assumed to start at 4~Myr in spectral synthesis models \citep[e.g. STARBURST99][]{leitherer2014}. However, which stellar mass goes into core-collapse SNe is not yet a settled question. The boundary mass for stars to explode as SNae is not yet well understood \citep{janka2025}. Moreover since massive stars in star clusters are those dominated most by stochastic IMF sampling, the latter needs to be taken into account for the majority of the eYSC population in M83. \cite{chevance2022pre} shows that SNe explosion are significantly delayed for low-mass clusters ($<$ $10^4$~M$_\odot$). In the meantime, stellar feedback in the form of photoionisation, radiation pressure, and stellar winds shape the star formation efficiency in the cloud, the IMF, the boundness of stars as well as the driving of HII region expansion \citep{ali2022, andersson2024, Menon2023, Lewis2023}.  This is in agreement with HII region studies conducted in M83 by \cite{della2022stellar}, and, in general, in local spiral galaxies \citep{barnes2021, Mcleod2021, pathak2025}. When SNe explode they will find an already processed medium and can therefore create large bubbles and clear the leftover gas of the natal clouds \citep[e.g., simulations][for observations]{grudic2022, andersson2024, sirressi2024}. The shorter timescales we observe in more massive clusters might be due to the likelihood of these clusters to host more massive stars and thus experience SNe feedback on shorter timescales. 

The emerging timescales reported in this work do not reflect the lifetime or destruction of the natal cloud. However, the timescales over which we see the PAH emission associated to clusters disappear (3--4 Myr), in agreement with many other works in the literature, indicate that pre-SN feedback is the main driver of PAH disruption in star-forming regions. Due to the complex nature of PAH and dust physics, simulations remains behind into shedding light on this point.

\section{Conclusions} \label{sec: conclusions}
We present the emerging star cluster population in the barred spiral galaxy M83 using new JWST NIRCam and archival HST observations. We detect eYSCs using NIR features tracing HII regions and PDRs, and combine it with the exposed oYSCs ($\leq10$ Myr) detected in the HST optical bands. The main goal of this paper is to estimate the emerging timescales necessary for gas clearing and how this varies as a function of cluster physical properties and diverse galactic environment. The main findings are:
\begin{itemize}
    \item Only a small fraction (20-30\%) of the eYSCs will probably survive as bound star clusters while the majority are consistent with stellar associations. This is expected by observations and simulations and confirms that most of star clusters may form within stellar associations.
    \item The NIR colors of the eYSCs and oYSCs indicate the presence of an evolutionary sequence, from embedded to exposed. We find that the morphological variations in the 3.3~$\mu$m PAH feature is consistent with an evolutionary sequence where eYSCIs evolve into eYSCIIs and optically visible oYSCs. This sequence is clear in the F300M-F335M (tracing the excess in the \PAHlambda\ feature) and the F115W-F187N (tracing the excess in Pa$\alpha$) colors which become increasingly bluer as clusters emerge. We also find that eYSCs are younger and more attenuated than the optical counterpart and that on average extinction is reduced with age.
    \item The mass distributions of the eYSCs and oYSCs are consistent with $-2$ power-law distributions. We observe variations across different galactic environments. The centre and the end of the bar stand out as dynamically active regions with high SFR which leads to the formation of massive star clusters. We then  exclude the central starburst region and analyze the disk cluster population in radial bins containing the same number of star clusters. We observe that the cluster mass function of both eYSCs and oYSCs shows similar distributions, suggesting that dynamical features as bar and spiral arms organize star formation in the galaxy but do not enhance it. 

    \item Over a timescale of 10~Myr, relative numbers of eYSCI, eYSCII and oYSCs suggest that the emergence sequence from embedded to fully exposed takes on average 6 Myr for the cluster population in M83. We find evidence that the initial embedded phase, prior to massive stars in the star cluster reaching the main-sequence and starting powering their HII region is very short. The phase over which eYSCI evolve into eYSCII and are no longer associated to \PAHlambda\ emission last about 4~Myr similarly derived by other JWST studies \citep[e.g.][]{rodriguez2023, linden2024, Whitmore_2023}. After this phase, the eYSCII remain associated to an HII region only for about 2~Myr, in agreement with pre-JWST studies \citep[e.g.][]{Messa2021,hannon22}.  We find evidence that the emergence sequence is shorter (5~Myr) for star clusters more massive than 5000~\msun, while it takes on average 7~Myr for a typical cluster of a few 1000~\msun\ to emerge. 
 
\end{itemize}
Overall the recovered timescales are longer than the nominal 4~Myr assumed for SNe explosions in stellar population models \citep{leitherer2014}. However, as showed by recent simulations \citep{grudic2022, Farias2024} and taking into account stochastic IMF sampling \citep{chevance2022pre}, SNe explosions in low mass clusters of a few 1000 \msun might be delayed. This implies that the emergence sequence is mainly driven by radiative and mechanical (in the form of stellar wind) pre-SN feedback and that at the time of first SN explosions their energy and momentum is injected in a processed medium, thus driving the final clearing stages.

Follow-up analysis of eYSCs and oYSCs populations for the entire sample of galaxies within the FEAST program will be pivotal in investigating emerging timescales in different galaxies and how they vary across a larger range of various galactic environments.

\section{Acknowledgments}

The FEAST team thanks Brent Groves for providing access to MAPPING-III spectral evolutionary models. A.A acknowledges support from Vetenskapsr\aa det
 2021-05559. A.A and A.P. acknowledge support from RS 31003240. A.A. and H.F.V. acknowledges support from RS 31004975.
E.A. acknowledges support from the NASA Astrophysics Theory Program grant 80NSSC24K0935. GÖ acknowledges support from the Swedish National Space Agency (SNSA). ASMB acknowledges the support from the Royal Society University Research Fellowship URF/R1/191609.
KG is supported by the Australian Research Council through the Discovery Early Career Researcher Award (DECRA) Fellowship (project number DE220100766) funded by the Australian Government. KG is supported by the Australian Research Council Centre of Excellence for All Sky Astrophysics in 3 Dimensions (ASTRO~3D), through project number CE170100013. 
 E.S. is supported by the international Gemini Observatory, a program of NSF NOIRLab, which is managed by the Association of Universities for Research in Astronomy (AURA) under a cooperative agreement with the U.S. National Science Foundation, on behalf of the Gemini partnership of Argentina, Brazil, Canada, Chile, the Republic of Korea, and the United States of America. ADC acknowledges the support from the Royal Society University Research Fellowship URF/R1/191609.
This work is based in part on observations made with the NASA/ESA/CSA James Webb Space Telescope (JWST). The data were obtained from the Mikulski Archive for Space Telescopes (MAST) at the Space Telescope Science Institute (STScI), which is operated by the Association of Universities for Research in Astronomy, Inc., under NASA contract NAS 5-03127 for JWST. These observations are associated with program $\#$ 1783. Support for program $\#$ 1783 was provided by NASA through a grant from STScI. Support to MAST for these data is provided by the NASA Office of Space Science via grant NAG 5–7584 and by other grants and contracts.This work is also based on observations made with the NASA/ESA Hubble Space Telescope, and obtained from the Hubble Legacy Archive, which is a collaboration between STScI/NASA, the Space Telescope European Coordinating Facility (ST-ECF/ESA), and the Canadian Astronomy Data Centre (CADC/NRC/CSA). This research has made use of the NASA/IPAC Extragalactic Database (NED) which is operated by the Jet Propulsion Laboratory, California Institute of Technology, under contract with NASA.

\vspace{5mm}
\facilities{JWST (NIRCam), HST (WFC3)}

\software{astropy \citep{astropy:2013, astropy:2018, astropy:2022},  
          Cloudy \citep{Ferland13}, 
          Source Extractor \citep{SEP_paper2, SEP_paper}, SAOImageDs9 \citep{ds9_paper}, NumPy \citep{numpy}, Pandas \citep{mckinney-proc-scipy-2010}, Matplotlib \citep{Hunter:2007}, Seaborn \citep{Waskom2021} }

\appendix
\renewcommand\theHtable{Appendix.\thetable}
\setcounter{table}{0}
\renewcommand{\thetable}{A\arabic{table}}

\setcounter{figure}{0}
\renewcommand{\thefigure}{A\arabic{figure}}
\section{Source extraction} \label{app:A}

We report the {\tt SEP} parameters for the source extraction and the reference image in Table~\ref{tab: source_extr_par}. We refer the reader to Section ~\ref{sec: method} for guidance.

\begin{deluxetable*}{lccc}
\tabletypesize{\scriptsize}
\tablewidth{0pt} 
\tablecaption{Source extraction parameters for the extraction of the eYSCs from the continuum-subtracted F187N, F335M and F405N, the F200W-based eYSCs and the optical YSCs.   
}
\label{tab: source_extr_par}
\tablehead{
\colhead{ Source extraction parameter/sources} & \colhead{eYSCs}& \colhead{F200W-based eYSCS} & \colhead{optical YSCs}\\
} 
\startdata 
     \hline
     Reference frame & F200W & F200W & F547M and F555W   \\
     Size of background box [pixels] &  30 & 30 & 30 \\
      Filter width and height  [pixels]  & 3  & 1 & 1 \\
      Threshold pixel value  & 5 & 10 & 5  \\
      Minimum area of pixels & 15 & 10 & 10\\
      Number of thresholds for deblending &32 & 32 & 32 \\
      Minimum contrast ratio for deblending &  0.00005 & 0.0005 &  0.0005\\
      \hline
\enddata
\end{deluxetable*}

\begin{deluxetable*}{cccccccccccc}
    \tabletypesize{\scriptsize} 
\tablewidth{0pt} 
  \tablecaption{Example of table which will be included in the published draft. Photometry extracted in the F200W reference frame, photometric catalog, magnitude in AB system, fluxes in mJy.}
    \label{tab: eYSCs_phot_table}
   \tablehead{
\colhead{id} & \colhead{Class} & \colhead{x} & \colhead{y}&  \colhead{ra} & \colhead{dec}  & \colhead{...} & \colhead{...} & \colhead{...} & \colhead{...} & \colhead{...} & }
\startdata 
    \hline
 1 &  \eYSCI & 7005.7389 &	3030.30680& 204.221908 &  $-$29.883107  &... &	... &	...&	...&	... & ...\\
       \hline
\enddata
\end{deluxetable*}

\section{Photometric catalogs}
We include here the photometric catalogues used in this paper. We refer the reader to Sections~\ref{sec: method} for the content and selection applied in this work.

      \begin{deluxetable*}{ccccccccccccc}
    \tabletypesize{\scriptsize} 
\tablewidth{0pt} 
  \tablecaption{photometric catalog, magnitude in AB system, fluxes in mJy.
    \label{tab: YSCs_phot_table}}
   \tablehead{
\colhead{id} & \colhead{Class} & \colhead{x} & \colhead{y}&  \colhead{ra} & \colhead{dec}  & \colhead{...} & \colhead{...} & \colhead{...} & \colhead{...} & \colhead{...} & }
\startdata  
    \hline
      & optical YSCs  &   6307.350364	 & 4668.986586	 & 204.266283 &	$-$29.879955 & &... &	... &	...&	...&	... & ...\\
       \hline
\enddata
\end{deluxetable*}

\section{Recovered oYSC physical properties over the entire M83 disk observed with HST}

We include here similar plots as presented in Figure~\ref{fig: distributions} of the main text. However, we include all the optical YSCs younger than 10 Myr extracted in the HST footprints, which cover an area 3 times larger than the JWST mosaic. Overall the extended oYSCs population shows similar properties as the sub-sample contained within the JWST FoV.
We refer the reader to Section~\ref{sec: results} of the main text.
\begin{figure*}
    \centering
    \includegraphics[width=0.9\linewidth]{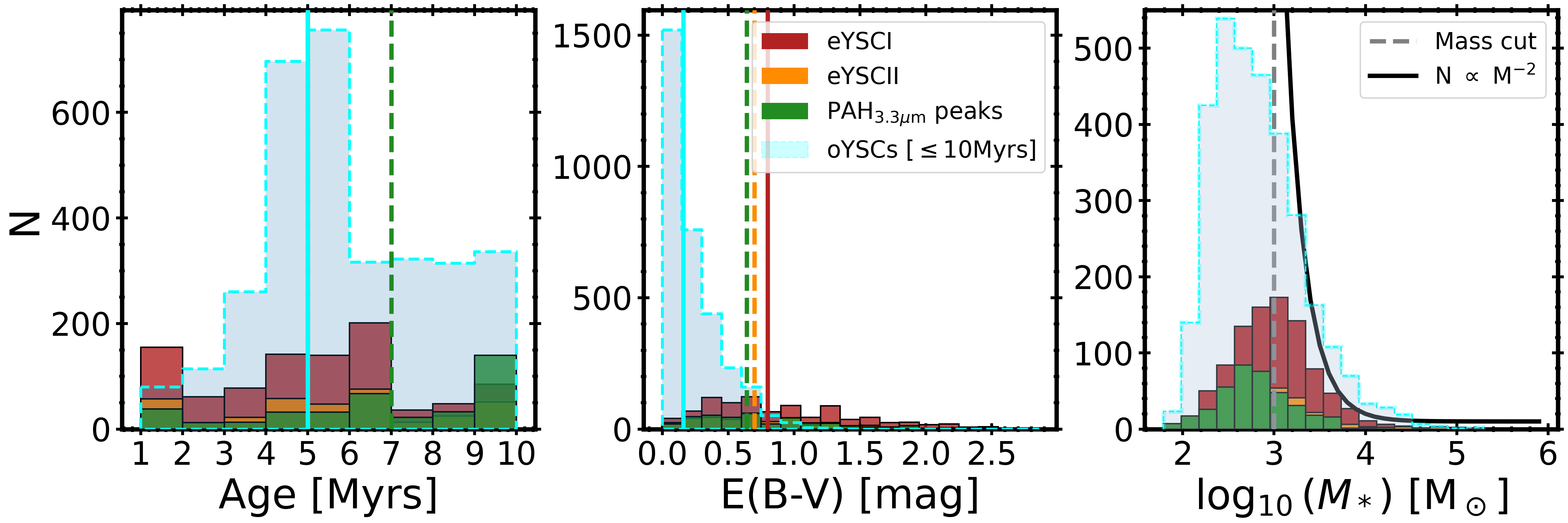}
    \caption{The stellar age (left), stellar attenuation (middle) and stellar mass distribution (right) for the different classes, \eYSCI\ (red), \eYSCII\ (orange), and \PAHclass\ (green) including the oYSCs ($\leq$ 10 Myr; cyan) for the HST FoV. The vertical lines correspond to the median values of \eYSCI\ (red), \eYSCII\ (orange), \PAHclass\ (green), and the YSCS ($\leq$ 10 Myr; cyan). The mass distribution (left) also includes a grey dashed line corresponding to the mass cut at $1\cdot10^3$ M$_\odot$ and the black curve to the 10-logarithm of a cluster mass function (CMF) with slope -2. For the stellar age (left panel) the \eYSCI\ median overlaps with the \eYSCII\ and  YSCs ($\leq$ 10 Myr). \label{fig: distributions_opticalfullfov}}

\end{figure*}

\bibliographystyle{aasjournal}



\end{document}